\documentclass[useAMS,usenatbib]{mn2e}

\usepackage{psfig}
\usepackage{epsfig}
\usepackage[varg]{txfonts}

\def\ltsima{$\; \buildrel < \over \sim \;$}
\def\lsim{\lower.5ex\hbox{\ltsima}}
\def\gtsima{$\; \buildrel > \over \sim \;$}
\def\gsim{\lower.5ex\hbox{\gtsima}}
\def\be{\begin{equation}}
\def\ee{\end{equation}}
\def\mec2{m_{\rm e}c^{2}}

\def\no{\noindent}

\renewcommand{\vec}[1]{{\bf #1}}

\begin{document}

\title[Disc emission from recoiling black holes]{Black hole mergers: the first light}
\author[Rossi et al.] {Elena~M. Rossi$^{1}$, G. Lodato$^{2,3}$, 
P.~J. Armitage$^{4,5}$, J.~E. Pringle$^{6,3}$ and A.~R. King$^{3}$\\
$^1$Racah Institute of Physics, The Hebrew University, Jerusalem 91904, Israel \\
$^2$Universit\`a degli Studi di Milano, Dipartimento di Fisica, via Celoria 16, I-20133 Milano, Italy \\
$^3$Department of Physics and Astronomy, University of Leicester, Leicester, LE1 7RH, UK\\
$^4$JILA, 440 UCB, University of Colorado, Boulder, CO 80309-0440, USA \\
$^5$Department of Astrophysical and Planetary Sciences, University of Colorado, Boulder, USA \\
$^6$Institute of Astronomy, Madingley Road, Cambridge, CB3 0HA, UK\\
\tt e-mail:  emr@phys.huji.ac.il (EMR)}

\maketitle
  
\begin{abstract}
The coalescence of supermassive black hole binaries occurs via the
emission of gravitational waves, that can impart a substantial recoil
to the merged black hole.  We consider the energy dissipation, that
results if the recoiling black hole is surrounded by a thin
circumbinary disc.  Our results differ significantly from those of
previous investigations.  We show analytically that the dominant
source of energy is often potential energy, released as gas in the
outer disc attempts to circularize at smaller radii.  Thus,
dimensional estimates, that include only the kinetic energy gained by
the disc gas, underestimate the real energy loss.  This underestimate
can exceed an order of magnitude, if the recoil is directed close to
the disc plane.  We use three dimensional Smooth Particle
Hydrodynamics (SPH) simulations and two dimensional finite difference
simulations to verify our analytic estimates.  We also compute the
bolometric light curve, which is found to vary strongly depending upon
the kick angle.  A prompt emission signature due to this mechanism may
be observable for low mass $(10^6 \ M_\odot)$ black holes whose recoil
velocities exceed $\sim 10^3 \ {\rm km \ s}^{-1}$. Emission at earlier
times can mainly result from the response of the disc to the loss of
mass, as the black holes merge. We derive analytically the condition
for this to happen.
\end{abstract}

\begin{keywords}
black hole physics --- accretion, accretion discs --- hydrodynamics
\end{keywords}

\section{Introduction}

 Supermassive black hole binaries are predicted by hierarchical galaxy
formation models and are a likely consequence of observed galaxy
mergers.  However, only a handful of these binaries have been 
directly observed \citep{rodriguez06} and their dynamical evolution is
still uncertain.  If binaries coalesce on a time-scale shorter
than the age of the universe, mergers can be an important ingredient
in the evolution and growth of supermassive black holes. Mergers 
also emit low frequency gravitational waves 
whose detection is one of the prime goals 
of proposed experiments such as the {\em Laser Interferometer 
Space Antenna} (LISA). What remains uncertain is whether detectable electromagnetic 
emission occurs either prior to or in the immediate aftermath of the 
coalescence. 

Observations and numerical simulations strongly suggest that 
mergers of gas-rich galaxies result in the inflow of gas into the 
nuclear region, where it is likely to co-exist with a newly 
formed black hole binary \citep{escala05,dotti07,callegari09}. 
The gas may have an active role and mediate and speed up the coalescence
\citep{ivanov99,natarajan02,cuadra09,lodato09}. However -- and irrespective 
of whether the gas is {\em dynamically} important -- if gas is present 
then perturbations to the
gas during the merger may give rise to observable  electromagnetic signals that can
either precede or follow the gravitational wave signal.  Observations
of these signals can give us evidence for the existence of mergers, 
improve upon LISA's limited ability to spatially localise sources, and
teach us about the astrophysical phenomena involved in the process.

 General relativity predicts that gravitational waves emitted by the
  shrinking binary during the final coalescence carry away a non-zero
  net linear momentum, so that the centre of mass of the merged black holes 
  recoils \citep{peres62,baker07}. A number of authors have used 
  semi-analytic methods or N-body simulations to estimate the 
  influence of the recoil on a surrounding thin disc \citep{lippai08,shields08,schnittman08}.  
  Hydrodynamic simulations for a thick disc -- a configuration that we do not consider here --
  have also been performed by \cite{megevand09}.
  
  Here we revisit the analytic problem (emphasizing the importance 
  of different physical effects from those discussed previously), 
  and perform numerical simulations to investigate how a
 thin, non self-gravitating disc reacts to the birth of a central
 black hole via merger.  
 The initial disc geometry that we consider is based upon that discussed
by \cite{natarajan02} and \cite{milos05}. We assume that, at a time
significantly prior to the final coalescence (when the binary
semi-major axis was $> 10^2 $ Schwarzschild radii), the binary was
surrounded by a geometrically thin disc whose plane coincided
with that of the binary. While the binary remains in tidal contact
with this disc, energy and angular momentum can be exchanged between
the two. This exchange provides an additional source of heat for the
inner disc \citep{lodato09}, and alters its long term viscous evolution
\citep{pringle91}. Immediately prior to the coalescence, however,
the rapid decay of the binary orbit due to gravitational wave
emission leads to a decoupling of the binary from the inner disc.
As a result, it is a reasonable first approximation to model the
effect of a kick on an initially unperturbed, axisymmetric
circumbinary disc. We also ignore any effects associated with
gas that has survived in circumprimary or circumsecondary
discs up until near the moment of coalescence. Such gas,
if present, may produce observable signatures \citep{natarajan02,
lodato09,chang09}, but it is decoupled dynamically from
the gas in the circumbinary disc. The effect of the recoil
on the circumbinary disc can therefore be considered
independently from the fate of gas bound to either of the
individual black holes.

 In this paper, our main focus is on 
 the effect of the black hole recoil rather than the almost instantaneous 
 mass loss that also accompanies the merger \citep{milos05}.  We study the
 consequence of velocity perturbations in both the linear (kick
 velocity smaller than circular velocity) and non linear
 regimes.  We calculate the magnitude of energy dissipation   
 and assess the dependence of this potentially 
 observable quantity on the recoil geometry. In particular, we focus
 on off-plane kicks, which are expected to be the most common.

The paper is organised as follows. In Section~\ref{sec:ana} we
analytically investigate the properties of the disc after the kick
and the dissipation of the extra energy in the innermost region.
These estimates serve as guidance for our numerical
simulations, described in Section~\ref{sec:sim}.
 In Section~\ref{sec:results} we present our numerical results,
which we discuss in Section~\ref{sec:discussion}.
Finally, in Section \ref{sec:conclusion} we draw our conclusions.

%------------------------------------------------------------------------------------------------------------
\section{Analytic estimates}
\label{sec:ana}

\subsection{Disc topology: bound and unbound regions}
\label{sec:regions}

\begin{figure}
\psfig{figure=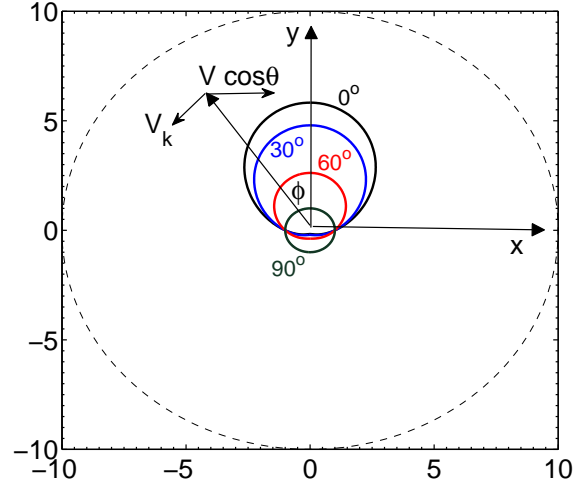,width=0.48\textwidth}
\caption[]{Face-on view of the disc showing the the Keplerian velocity
$V_{\rm k}$ and the projection of the kick velocity onto the disc
plane $V \cos{\theta}$. The angle between the positive $y$-axis and the
particle's position is $\phi$.  The closed solid lines mark the border
between the bound and unbound regions for kick angles between $0^\circ$ (a kick 
directed into the disc plane) 
and $90^\circ$. As $\theta$ approaches 90 degrees, the bound region becomes 
increasingly azimuthally symmetric.  }
\label{fig:topo}
\end{figure}

We first consider the reaction to the kick of 
particles at radius $R$. Initially, the particles rotate anti-clockwise
around the black hole in circular orbits with Keplerian
velocity $\vec{V_{\rm k}} = \sqrt{GM/R}\; \hat{\phi}$, where $\phi$ is
the angle between the positive $y$-axis and the particle radius $R$,
$M$ is the mass of the merged black hole and $G$ is the gravitational
constant (see Fig.~\ref{fig:topo}). The angular velocity is 
$\Omega = \sqrt{GM/R^3}$. 

In the frame of the moving central object the recoil results in 
each particle receiving an an additional velocity $\vec{V}$. 
The kick velocity direction makes an angle $\theta$
with the disc plane. For simplicity, we assume that the projection
onto the disc plane, $V \cos \theta$, is in the direction of the
positive $x$-axis (see Fig.~\ref{fig:topo}). In cylindrical 
coordinates the three components of the initial particle 
velocity are then $V_{\rm R} = -V
\cos{\theta} \sin{\phi} $, $V_{\phi} = V_{\rm k}-V \cos{\theta}
\cos{\phi}  $ and $V_{\rm z} = V \sin{\theta}$.

If the disc extends to sufficiently large radii, there is a radius
$R_{\rm v}$ at which $V = V_{\rm k}$. In the following, it is
convenient to adopt the dimensionless radial coordinate $r= R/R_{\rm
v}=\left(V/V_{\rm k}\right)^2 $. The specific energy of a particle
then reads

\be
\epsilon= -\frac{1}{2} V_{\rm k}^2 \left(1+2 \sqrt{r} \cos{\theta}\cos{\phi}-r \right).
\label{eq:E}
\ee

\no A particle can escape from the system ($\epsilon>0$)
 or it can remain bound to it ($\epsilon<0$), depending on its
 distance from the hole and on the direction of the kick velocity
 relative to its original Keplerian velocity. In particular at small radii,

\be
r< r_{\rm b}=\left(-\cos{\theta}+\sqrt{\cos{\theta}^2+1}\right)^2,
\label{eq:rb}
\ee

\no
 all particles within a given annulus are bound, while for

\be
r> r_{\rm ub}=\left(\cos{\theta}+\sqrt{\cos{\theta}^2+1}\right)^2,
\label{eq:rub}
\ee

\no all particles are unbound. The dimensionless radii $r_{\rm b}=R_{\rm
b}/R_{\rm v}$ and $r_{\rm ub}=R_{\rm ub}/R_{\rm v}$ correspond to the
actual radii $R_{\rm b}$ and $R_{\rm ub}$, respectively.  In the
radial range between $r_{\rm b}$ and $r_{\rm ub}$, particles at radius
$r$ are bound only within a limited azimuthal range $-\phi_{\rm
b}<\phi <\phi_{\rm b}$ (with $0\le \phi_{\rm b} \le \pi$), where

\be
\cos{\phi_{\rm b}}= \frac{r-1}{2\sqrt{r}\cos{\theta}}.
\label{eq:phib}
\ee

\no
In Fig.~\ref{fig:topo} we show the boundaries of the bound region
for different kick angles. As $\theta$ increases, the bound region becomes increasingly azimuthally symmetric,  
while the radial extent of the bound region between $r_{\rm b}$ and $r_{\rm ub}$ decreases. 
For a perpendicular kick $r_{\rm b}=r_{\rm ub}=1$.

%%%%%%%%%%%%%%%%%%%%%%%%%%%%%%%%%%%%%%%%%%%%%%%%%%%%%%%%%%%%%%%%%%%                                  
%%%%%%%%%%%%%%%%%%%%%%%%%%%%%%%%%%%%%%%%%%%%%%%%%%%%%%%%%%%%%%%%%%% 

In our analytic estimates we consider only  the portion of the disc that is still bound after the kick,
since matter that is unbound by the recoil simply leaves the system and it is not energetically important.

\subsection{Angular momentum}
\label{sec:ang_mom}
In the bound portion of the disc (where $\epsilon <0$), particles
are excited on elliptical orbits with eccentricity $e^2 = 1 + 2 j^2
\epsilon/(G M)^{2}$, and with specific angular momentum

\be
\frac{\vec{j}}{J_{\rm k}} =  \left(1-\sqrt{r} \cos{\theta} \cos{\phi} \right)\hat{z} - \sqrt{r} \sin{\theta}\; \hat{\phi},
\label{eq:h}
\ee

\no
where $J_{\rm k}=R V_{\rm k}$ is the Keplerian specific angular momentum for that radius. 
In general $\vec{j}\neq \vec{J_{\rm k}}$, and as a consequence rearrangement of gas 
must occur after the recoil as the gas seeks a new minimum energy configuration. 
If a particle conserves its angular momentum then it must circularize at a radius 
$r_{\rm c,p} = R_{\rm c,p} /R_{\rm V}$ given by

\begin{equation}
 \frac{r_{\rm c,p}}{r}= \left( \frac{j}{J_k} \right)^2 = \left( 1 - \sqrt{r} \cos \theta \cos \phi \right)^2 
 + r \sin^2 \theta.
\label{eq:rcp} 
\end{equation}
\no
To understand the behaviour of the disc after the kick, we calculate
the mean circularization radius for a ring, $r_{\rm c} = R_{\rm c} /R_{\rm V}$,

\begin{eqnarray}
\frac{r_{\rm c}}{r} = \frac{1}{2 \phi_{\rm b}} \int_{-\phi_{\rm b}}^{\phi_{\rm b}}\left(\frac{j}{J_{\rm k}}\right)^{2} \;d\phi = 
(1+r\sin^2{\theta})+ \nonumber \\ r \cos^2{\theta} \left(\frac{1}{2}+\frac{\sin{2 \phi_{\rm b}}}{2 \phi_{\rm b}}\right)-2 \sqrt{r}
\cos{\theta} \frac{\sin{\phi_{\rm b}}}{\phi_{\rm b}},
\label{eq:rc}
\end{eqnarray}

\begin{figure}
\psfig{figure=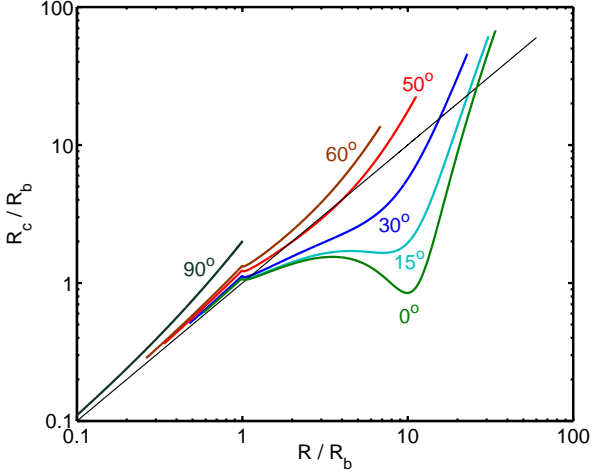,width=0.48\textwidth}
\caption[]{The mean circularisation radius (equation~\ref{eq:rc}) in units of
$R_{\rm b}$ (equation~\ref{eq:rb}) as a function of the initial location
$R$ in the same units. The black solid thin line ($R=R_{\rm c}$) divides the region where
outward spreading occurs (upper part $R_{\rm c}> R$) from the region
where accretion occurs ($R_{\rm c}<R$).  }
\label{fig:rc}
\end{figure}

\no
where the integration limits are such that $\phi_{\rm b}=\pi$, for $r\le r_{\rm b}$.
Fig.~\ref{fig:rc} shows that
while matter at $R<R_{\rm b}$ circularizes close to its
original location for any kick angle, $R_{\rm c}$ depends strongly on
the kick angle for $R>R_{\rm b}$. For high latitude kicks ($\theta
\gsim 50^{\circ}$), the behaviour outside and inside $R_{\rm b}$ is
similar. For smaller inclination angles an increasing amount of matter 
circularizes at smaller radii, and only a small fraction of matter expands 
toward larger radii. This implies that there must be a net release of potential energy.

For out-of-the plane kicks, the recoiling matter acquires in general an 
{\it azimuthal} component in its new angular momentum (equation~\ref{eq:h}). 
Its magnitude varies with radius, so we expect the disc to be warped initially. 
After the extra energy associated with the warp is dissipated, the disc must lie in
a plane, however tilted with respect to its original orientation.

%%%%%%%%%%%%%%%%%%%%%%%%%%%%%%%%%%%%%%%%%%%%%%%%%%%%%%%%%%%%%%%%%%%                                  
%%%%%%%%%%%%%%%%%%%%%%%%%%%%%%%%%%%%%%%%%%%%%%%%%%%%%%%%%%%%%%%%%%% 

\subsection{Energy dissipation}
In the previous section, we showed that particles that remain bound after the kick do not, in general, have 
the same specific angular momentum as they had prior to the kick. 
Initially their orbits are eccentric. In a fluid disc, orbit crossing 
leads to energy dissipation and circularization of the 
disc into a new circular equilibrium configuration.

We estimate the total energy available for dissipation by assuming 
that each particle that remains bound circularizes at the radius 
appropriate to its post-kick angular momentum. We write this as,
\begin{equation}
 E_{\rm diss} = E - E_{\rm c},
\label{eq:ediss} 
\end{equation}
where $E$ is the initial post-kick energy of the bound disc (including the 
gain or loss of kinetic energy resulting from the kick) and $E_{\rm c}$ 
is its energy after circularization. The initial energy 
is obtained by integrating the specific energy (equation~\ref{eq:E}) 
over the bound region of the disc,
\begin{equation}
 E = \int_{R_{\rm in}}^{R_{\rm ub}} \int_{-\phi_{\rm b}}^{\phi_{\rm b}} 
 \epsilon \; \Sigma \, R \, dR \, {\rm d} \phi.
\end{equation}
Here $\Sigma$ is the disc surface density, $R_{\rm in}$ is the inner radius of the 
disc, and it is to be understood that 
the limits on the angular integral are such that $\phi_{\rm b} = \pi$ 
when $R \leq R_{\rm b}$. 

After circularization the total energy of the gas within the disc is 
\begin{equation}
 E_{\rm c} = \int_{R_{\rm in}}^{R_{\rm ub}} \int_{-\phi_{\rm b}}^{\phi_{\rm b}} 
 -\frac{1}{2} V_{\rm k}^2 \left( \frac{r}{r_{\rm c,p}} \right) \; \Sigma \, R \,dR \,{\rm d} \phi, 
\end{equation}
where $(r/r_{\rm c,p})$ is given by equation~(\ref{eq:rcp}). 
The integral over $\phi$ is analytic and can readily be evaluated using a symbolic algebra
package. However the simplest general form we have found is extremely lengthy, and we do not write it
here.

\begin{figure}
\psfig{figure=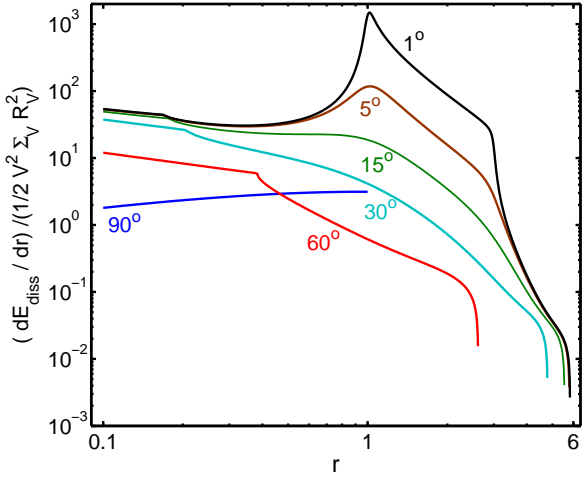,width=0.48\textwidth}
 \caption[]{Differential energy to be dissipated is plotted as a function of radius 
(equation~\ref{eq:de}). The energy is expressed in units of $(1/2) V^2 \Sigma_V R^2_{\rm V}$, while the radius is in units of $R_{\rm V}$.
 Different curves correspond to different kick angles as labelled. The surface density profile is 
$\Sigma \propto r^{-3/2}$.}
\label{fig:de}
\end{figure}

\begin{figure}
\psfig{figure=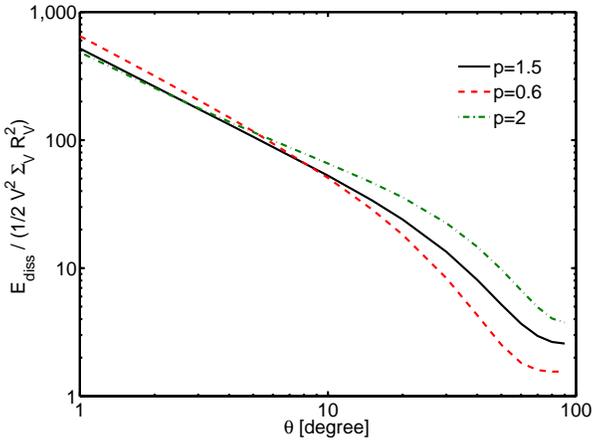,width=0.48\textwidth}
 \caption[]{Total energy available to be released (equation~\ref{eq:ediss}) is plotted 
as a function of the kick angle for different power-law indexes for the surface density distribution.
 The energy is expressed in units of $(1/2) V^2 \Sigma_V R^2_{\rm V}$.}
\label{fig:ediss}
\end{figure}

To assess which parts of the disc contribute the most to $E_{\rm diss}$ we now 
specialize to a disc with a power-law surface density profile $\Sigma = \Sigma_{\rm V} r^{-p}$, 
where $\Sigma_{\rm V} = \Sigma (r=1)$. Dimensional arguments\footnote{For example, 
\cite{schnittman08} estimate the total energy release to be $\approx 4 (V^2 / 2) \Sigma_{\rm V} R_{\rm V}^2.$} 
suggest that the change in kinetic energy 
is of the order of $(V^2 / 2) \Sigma_{\rm V} R_{\rm V}^2$.  
Hence we write the differential contribution to the total energy 
release as,
\begin{equation}
\frac{{\rm d} E_{\rm diss}}{{\rm d} r}= \frac{V^2}{2} \Sigma_{\rm V} R_{\rm V}^2 \; g(r),
\label{eq:de}
\end{equation}
where,
\begin{eqnarray}
g(r) &=&  2 \phi_{\rm b} \; r^{-p} \;  \left[ r-2\sqrt{r}\cos{\theta}  \frac{\sin{\phi_{\rm b}}}{\phi_{\rm b}} -1+ {\cal
I}(r)\right], \\
{\cal I}(r) &=&   \frac{1}{2 \phi_{\rm b}} \int_{-\phi_{\rm b}}^{\phi_{\rm b}}\left(\frac{j}{J_{\rm k}}\right)^{-2} \;d\phi.
\end{eqnarray}
The function $g(r)$ is a measure of how well the simple estimate 
$(1/2) V^2 \Sigma_V R_V^2$ captures the expected total energy release. 
This function is plotted for different values of the kick angle (and $p=1.5$) in Fig.~\ref{fig:de}. 
As $\theta$ decreases, the behaviour around $r=1$ is increasingly 
dominated by ${\cal I}$ and it develops a sharp peak at $r=1$. 
For $\theta =0^\circ$, the right-hand side of equation~(\ref{eq:de}) diverges as $|r-1|^{-1}$.
For $\theta \neq 0^\circ$ we integrate over radius to obtain the 
final estimate of $E_{\rm diss}$. The dependence of $E_{\rm diss}$ 
on kick angle is shown in Fig.~\ref{fig:ediss}. We find that 
regardless of the surface density distribution, $E_{\rm diss}$ is
almost 3 orders of magnitude larger for a kick close to the disc plane 
than for a perpendicular kick. Moreover the assumption made in previous 
papers that $(1/2) V^2 \Sigma_V R^2_{\rm V}$ would fairly measure $E_{\rm diss}$ 
proves to be a substantial underestimate of the energy released for all kick angles 
$\theta < 50^{\circ}$.
  
%%%%%%%%%%%%%%%%%%%%%%%%%%%%%%%%%%%%%%%%%%%%%%%%%%%%%%%%%%%%%%%%%%%                                  
%%%%%%%%%%%%%%%%%%%%%%%%%%%%%%%%%%%%%%%%%%%%%%%%%%%%%%%%%%%%%%%%%%% 

%%%%%%%%%%%%%%%%%%%%%%%%%%%%%%%%%%%%%%%%%%%%%%%%%%%%%%%%%%%%%%%%%%%                                  
%%%%%%%%%%%%%%%%%%%%%%%%%%%%%%%%%%%%%%%%%%%%%%%%%%%%%%%%%%%%%%%%%%% 

\subsection{Contributions to $E_{\rm diss}$}
The energy available as the disc adjusts to a new equilibrium configuration 
can be considered to have two sources. First, there is an immediate change 
to the kinetic energy of particles because of the recoil. When integrated over the 
bound region the total energy change from this effect is positive for $\theta > 4^{\circ}$. 
Second, there is the release of potential energy that 
is liberated when gas in the disc loses angular momentum and circularizes 
at smaller radii. In the remainder of the paper we will refer to the release 
of potential energy on a short time scale as ``accretion energy". This should 
not be confused with the (much slower) energy release due to viscous disc 
accretion (described as ``accretion energy" by Schnittman \& Krolik 2008), which 
we do not model. 

To quantify how the relative importance of the two energy sources  
varies with radius and with kick angle we rewrite the post-kick specific energy as 
\begin{equation}
 \epsilon = -\frac{1}{2} \, V_{\rm k}^2 -\frac{1}{2} V_{\rm k}^2  \left( 2 \sqrt{r} \cos \theta \cos \phi - r \right) =  -\frac{1}{2} \, V_{\rm k}^2 + \epsilon_{\rm kick},
\end{equation}
and define the prompt change to the energy of the disc as,
\begin{equation}
 E_{\rm kick} = \int_{R_{\rm in}}^{R_{\rm ub}} \int_{-\phi_{\rm b}}^{\phi_{\rm b}} 
 \epsilon_{\rm kick} \; \Sigma \, R \, dR \, {\rm d} \phi.
\end{equation}
We note immediately that if $\theta = 90^\circ$ then $\epsilon_{\rm kick}$
 is simply equal to $V^2 / 2$. Integration over the bound region of the 
disc then yields $E_{\rm kick} = (1/2) M_{\rm disc} V^2$, where $M_{\rm disc}$ 
is the total mass within the bound region.

\begin{figure}
\psfig{figure=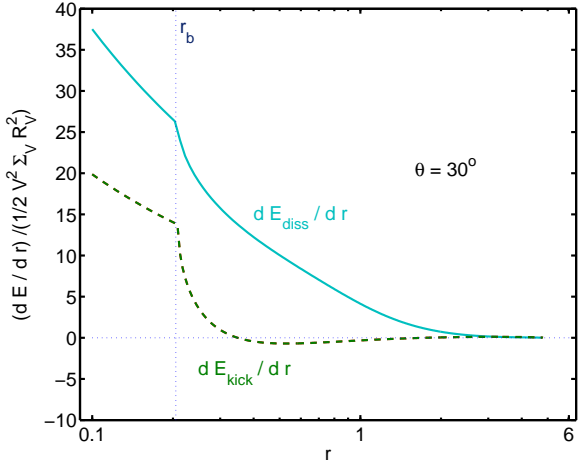,width=0.48\textwidth}
\caption[]{Differential energy available for dissipation 
(solid line) and change in total kinetic energy imparted directly by the kick (dashed line) 
is plotted as a function of radius, for $\theta = 30^{\circ}$. As previously,
energy is in units of $(1/2) V^2 \Sigma_V R^2_{\rm V}$, the radius is in units of $R_{\rm V}$, 
and $\Sigma \propto r^{-3/2}$.}
\label{fig:diss_kick}
\end{figure}

For arbitrary kick angles one can readily show that the contribution to $E_{\rm kick}$ 
for $R \leq R_{\rm b}$ is given by the same expression,
\begin{equation} 
 \Delta E_{\rm kick} = \frac{1}{2} \Delta M_{\rm disc} V^2,
\end{equation}
but where $\Delta M_{\rm disc}$ is now the mass within the radius where the entire 
annulus remains bound to the hole.  At larger radii, however, this is no longer true. For 
$R_{\rm b} < R < R_{\rm ub}$ the differential contribution to the prompt energy 
change ${\rm d} E_{\rm kick} / {\rm d} R$ can be either positive or negative, depending 
upon the radius and upon the specific kick angle. This is illustrated in 
Fig.~\ref{fig:diss_kick} for the case of a kick at $\theta = 30^{\circ}$. 
When the integration is extended across the entirety of the bound disc the result is that 
 $E_{\rm kick} \neq (1/2) M_{\rm disc} V^2$ for $\theta \neq 90^\circ$.

Finally we can compare the relative importance of the prompt kinetic energy 
to the part of $E_{\rm diss}$ that arises from potential energy release. 
In general, we find that for $R > R_{\rm b}$ the gain in kinetic energy is 
either negative or (when positive) negligible compared to the accretion 
energy, whereas for $R \leq R_{\rm b}$ it can be substantial (Fig.~\ref{fig:diss_kick}). 
As we will discuss subsequently the physical size of $R_{\rm b}$ is often quite 
large, so for a small disc that does not extend out to $R_{\rm b}$ it may be a reasonable 
approximation to assume that the disc receives an injection of kinetic 
energy that is subsequently dissipated. For a large disc, on the other 
hand, any kick that is not nearly perpendicular to the disc plane produces 
a wholesale rearrangement of the gas. The 
total energy dissipation in this regime is dominated by accretion energy, 
and detailed hydrodynamic simulations are needed to assess accurately the 
magnitude and time scale of energy release.

When we consider the energies integrated over the whole bound disc,
we find that as $\theta$ decreases the 
fraction of $E_{\rm diss}$ that is due to release of potential energy increases.
This is because the recoil creates larger gradients in the specific energy and angular momentum 
of the particles. Two effects result. First, the region between $r_{\rm b}$ and $r_{\rm ub}$ grows and, second, 
within that region more particles have a new lower specific angular momentum (see Fig.~\ref{fig:rc}).
 Therefore, hydrodynamic simulations are needed, especially, for kicks grazing the plane
and above all, for in-plane kicks, where the analytic estimate of $E_{\rm diss}$ fails.

%%%%%%%%%%%%%%%%%%%%%%%%%%%%%%%%%%%%%%%%%%%%%%%%%%%%%%%%%%%%%%%%%%%                                  
%%%%%%%%%%%%%%%%%%%%%%%%%%%%%%%%%%%%%%%%%%%%%%%%%%%%%%%%%%%%%%%%%%% 

\subsection{Effect of mass variation of the post-remnant black hole}
\label{sec:analytic_mass_loss}

Not only momentum, but also energy is carried away by
the gravitational waves that cause the final merger of a black hole 
binary. This results in a net mass loss for the system and the final black hole born
by merger has a smaller mass than the sum of the masses of the
two progenitors.  This instantaneous mass loss of the central
attractor has an effect on the surrounding gas \citep{milos05,schnittman08,oneill09,megevand09}.
We show here that this effect deposits
substantial energy {\em only} within the innermost $\sim 10^3$
Schwarzschild radii for $V > 100 \ {\rm km \ s}^{-1}$.

The argument is as follows. We first estimate the energy dissipated in the gas as the disc
recovers a circular equilibrium configuration around a lighter central
object, neglecting the effect of recoil. The effect of the  mass loss $\delta M$ is to put particles in eccentric orbits,
since it instantaneously decreases their binding energy. The new specific
energy of a ring initially at radius $R$ is $\epsilon_{\rm i}= -GM/(2 R)\,
(1- 2\, \delta M/M)$. In the circularization process, the particles
settle on a circular orbit at a {\it larger} radius $R_{\rm c}$,
given by angular momentum conservation: $R_{\rm c} = R /(1-\delta
M/M)$. At this final radius the specific energy is $\epsilon_{\rm f}= -GM/(2
R)\, (1-\delta M/M)^{2}$.  Thus, the induced epicyclic motions
dissipate an energy 

\be \Delta \epsilon_{\rm m} = \epsilon_{\rm i}-\epsilon_{\rm
f}=\frac{1}{2} V_{\rm k}^2 \left(\frac{\delta M}{M} \right)^2.
\label{eq:dm}
\ee

\begin{figure}
\psfig{figure=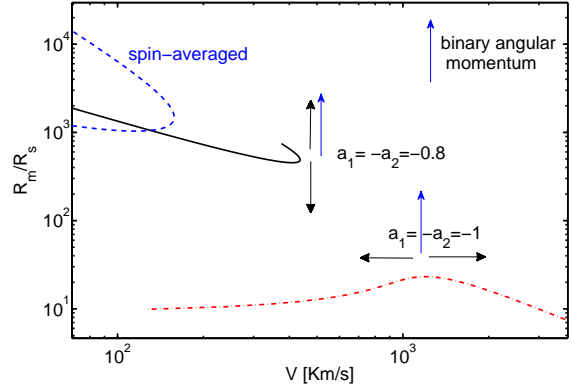,width=0.48\textwidth}
\caption[]{Radius within which $\Delta \epsilon_{\rm m}$ dominates over
$V^2/2$ as a function of the kick velocity $V$.  The
dashed line is an average over the spins, assuming isotropic
distribution of spin orientation (it amounts to the curve for the kick
velocities due only to different mass ratios for the BHs). The solid
line assumes equal magnitude spins, aligned with the orbital angular
momentum vector but pointing in opposite directions. Finally, the
dot-dashed line is for equal magnitude spins, in the direction
perpendicular to the orbital angular momentum and anti-aligned.  }
\label{fig:rm}
\end{figure}

We now assess the relative importance at a given radius of the effects
of the mass loss compared to that of the recoil. Equation~(\ref{eq:dm}) shows that
$\Delta \epsilon_{\rm m} \propto R^{-1}$ is a decreasing function of
radius. Since the contribution of mass loss turns out to be important only within $R_{\rm b}$, 
we can compare $\Delta \epsilon_{\rm m}$ with the energy imparted by the recoil $ \approx  V^2/2$. We find that there is a radius $R_{\rm m} < R_{\rm b}$ {\it
within} which $\Delta \epsilon_{\rm m}>V^2/2$,
\be
\frac{R_{\rm m}}{R_{\rm s}} = \frac{1}{2} \left(\frac{c}{V} \right)^2 \left(\frac{\delta M}{M}\right)^2,
\label{eq:rm}
\ee
where  $R_{\rm s} = 2GM/c^2$ is the Schwarzschild radius and $c$ is the speed
of light.  Both $\delta M/M$ and $V$ depend on the mass ratio and on the
magnitudes and directions of the spins of the merging black holes.  To evaluate
$R_{\rm m}$ we use the analytic expressions, calibrated against 
simulations, given by \cite{tichy08} and \cite{baker07}
respectively.  The resulting value of $R_{\rm m}/R_{\rm s}$ is shown in Fig.~\ref{fig:rm}, for three
different configurations of the black hole spins ($a_{1}$, $a_{2}$).
For some choices of the parameters, two different magnitudes for the mass loss are associated
with the same $V$, explaining the double valued behaviour of $R_{\rm
m}$\footnote{While $\delta M$ depends only on the total energy emitted
in gravitational waves, $V$ depends also on the degree of asymmetry 
of the emission. Therefore, two different configurations at merger can
release different total energy but carry away the same momentum.}.  The
figure shows that the larger the kick, the smaller the radius of
influence of the mass loss, approaching the innermost stable orbit for
the most extreme recoils. 

%%%%%%%%%%%%%%%%%%%%%%%%%%%%%%%%%%%%%%%%%%%%%%%%%%%%%%%%%%%%%%%%%%%                                  
%%%%%%%%%%%%%%%%%%%%%%%%%%%%%%%%%%%%%%%%%%%%%%%%%%%%%%%%%%%%%%%%%%% 

In our simulations, we verify that for our set-up the dissipation
from mass loss is indeed negligible at all times. We will show an example
 of how the lightcurve from the recoil compares with that of the mass loss for an
in-plane kick in Section \ref{sec:massloss}.

\section{Simulations}
\label{sec:sim}

Our analytic predictions are based upon a {\em particle} rather than 
a {\em fluid} model for the disc. Thus, we 
have run a set of simplified numerical hydrodynamic simulations, 
in order to check these predictions, and to obtain the explicit form of the energy 
dissipation rate within the disc as a function of time. 
We discuss first the special cases of perpendicular ($\theta=90^\circ$) 
and in-plane ($\theta=0^\circ$) kicks. The perpendicular 
case is exactly axisymmetric, while the in-plane case is 
also a two-dimensional problem (in $r$ and $\phi$) in the 
limit of an infinitesimally thin disc. 
Given their symmetry, these cases can be simulated efficiently using 
Eulerian finite difference methods. The general case of 
out-of-plane kicks and the $\theta=0^\circ$ case when the disc 
has a finite thickness are fully three-dimensional, and correspondingly 
more difficult. We treat these cases using the Lagrangian Smoothed 
Particle Hydrodynamics (SPH) method \citep{gingold77,benz90,monaghan92}. 
We compare the two numerical 
methods by simulating the $\theta=90^\circ$ case with both 
codes.

\subsection{Initial set-up}
\label{sec:set_up}

We investigate a thin disc, extending from $r=0.1$ to $r =10$, in order to
encompass the three regions described in Section \ref{sec:regions}.
The radial distribution of surface density and sound speed, $c_{\rm
s}$, are power-laws 

\begin{equation}
 \Sigma (r) \propto r^{-p} 
 \label{eq:cs_sig1}
 \end{equation}
 \be
 c_s (r) \propto r^{-3/4},
\label{eq:cs_sig2}
\end{equation}
with the normalization of the sound speed chosen such that the disc
aspect ratio $H/R = 0.05$ at the characteristic radius $R_V$. The
initial conditions of the simulations are isothermal with a fixed (in
time) but radially dependent sound speed. In the finite difference
runs the sound speed profile at later times remains fixed (in Eulerian
co-ordinates), whereas in the SPH run the internal energy per unit
mass of each {\em particle} remains constant (i.e. the sound speed is
fixed in a Lagrangian sense).  In either case the physical assumption
is that the energy dissipated as the disc readjusts to its new
equilibrium configuration is radiated away rapidly.  Since we do not
include the effects of the disc self-gravity, the surface density
normalization is arbitrary. We can freely scale our results to
represent a particular choice of the disc to black hole mass ratio $q
= M_d / M$.

The unperturbed --- pre-kick --- disc velocity field
is set to give centrifugal equilibrium in the black hole gravitational field (assumed to be Newtonian). 
The 
correction arising from radial pressure gradients within the disc is taken into account, even if it is
small for the thin discs that we consider. 
Working in the frame comoving with the black hole, 
the initial (perturbed) velocity field is a superposition of
Keplerian anticlockwise rotation with a kick of magnitude $V$, directed
at an angle $\theta$ from the disc. 

\subsection{Numerical implementation}

As mentioned above, we performed numerical simulations using two very
different schemes.  We have compared the corresponding results in
order to check that they are numerically reliable. In this section we
give an overview of both codes, and in particular note those features
that are most important for understanding our conclusions.

We model the effect of an in-plane or perpendicular kick using the
ZEUS code \citep{stone92}. ZEUS is a Eulerian finite-difference code.
For in-plane kicks
we use cylindrical polar co-ordinates $(r,\phi)$, assuming symmetry in
the $z$ direction. For perpendicular kicks we use spherical
polar co-ordinates $(r,\theta)$, assuming symmetry in $\phi$.  We
impose outflow boundary conditions at the inner and outer edges of the
grid. Outflow boundary conditions are implemented in ZEUS by setting
the gradient of all physical quantities to zero at the boundary. This
is exact only for supersonic outflow. Exploratory runs showed that for
in-plane kicks enough accretion occurs to bring subsonic gas into
contact with the inner boundary. This leads to unphysical
reflections. To ameliorate this, we additionally impose a near-vacuum
in the innermost active grid zone by sweeping away gas above some
small threshold on every time step.

SPH is a Lagrangian code to solve the hydrodynamics equations. Our
version of the code self-consistently includes the so-called $\nabla
h$ terms needed to ensure energy conservation \citep{price07}, where
$h$ is the smoothing length, which sets the effective resolution
length of the simulation. The smoothing length is adjusted adaptively
in the code, ensuring higher resolution in high density regions. For a
disc with surface density and sound speed profiles as in
equations~(\ref{eq:cs_sig1}) and (\ref{eq:cs_sig2}), with $p=3/2$, the
azimuthally averaged smoothing length $\left <h \right>\propto
H$. Therefore, this disc vertical structure is equally resolved at
each radius. We have run SPH simulations using different total number
of particles $N$ in order to check for convergence. We
have found that our results, and in particular the expected luminosity
arising from the kick, converge at high resolution, for $N> 4$
million. In the following we present only results corresponding to our
highest resolution simulations, which employ either 4 or 8 million
particles.

Both codes use artificial 
viscosity to capture shocks. In our simulations 
the stress tensor associated with artificial viscosity enters the momentum 
equation but is neglected in the energy equation (indeed, no 
energy equation is required in our isothermal runs). Nonetheless, 
we can compute the {\em implied} rate at which viscosity is 
dissipating kinetic energy in shocks and compressions. Use 
this to measure the spatial and temporal dependence of 
the energy dissipation caused by the kick. 

In ZEUS, in one spatial 
dimension (say $r$) the implied rate of change of the internal 
energy per unit volume $e$ is,
\begin{equation}
 \frac{ \partial e}{\partial t} = - Q \left( \frac{\Delta v_r}{\Delta r} \right),
\end{equation}
where $\Delta v_r = v_{r, i+1} - v_{r, i}$ is the change in velocity 
across one cell of width $\Delta r$ and,
\begin{equation}
 Q = \left\{ 
 \begin{array}{ll}
 C \rho \left( \Delta v_r \right)^2 & {\rm if \, \Delta v < 0} \\ 
 0 & {\rm otherwise}.
 \end{array}
 \right.
\label{eq:q_zeus}
\end{equation} 
The dimensionless constant $C$ controls the numerical width of shock 
fronts. We use $C=2$. Following the operator-split approach used 
throughout ZEUS we calculate the full heating rate by adding 
independent one-dimensional terms in $r$ and $\phi$ (for in-plane 
kicks) or $r$ and $\theta$ (for perpendicular kicks).

In the SPH code the fundamental quantity is the energy per unit mass $u$.
The implied rate of change of  $u$ due to artificial viscosity is  
\be
 \frac{\partial u}{\partial t} = - \frac{Q}{\rho} \nabla \cdot {\bf v},
\ee
where the divergence of the velocity is calculated over a smoothing length $h$
and the artificial viscosity term $Q$ is

\begin{equation}
 Q = \left\{ 
 \begin{array}{ll}
 \alpha c_{\rm s} h \nabla \cdot {\bf v}+ \beta h^2 \; (\nabla \cdot {\bf v})^2 & {\rm if \,\nabla \cdot {\bf v}  < 0} \\ 
 0 & {\rm otherwise}.
 \end{array}
 \right.
\label{eq:q_sph}
\end{equation}
We use the viscosity switch described 
in Morris \& Monaghan (1997), where the parameter $\beta =2 \alpha$ and 
$\alpha$ varies from  $1$ to $ 0.01$ so that
dissipation is enhanced when the particle enters the shock region and reduced otherwise. 

We note that the quadratic terms for artificial viscosity are essentially
equivalent in the two codes, although the SPH implementation is
{\em not} operator split.  To calculate the total luminosity that is
released (thereafter labelled ``implied $dE / dt$''), we sum the implied rate of energy dissipation
(equation~(\ref{eq:q_zeus}) or equation~(\ref{eq:q_sph})) over volume for ZEUS, and
over mass for SPH.

\section{Numerical results}
\label{sec:results}
In the following, results are presented in code units in 
which the unit length is taken to be 
\be
R_{\rm V}=GM/V^2
\ee

\no
and the unit time is the dynamical time 
\begin{equation}
 t_{\rm V} = \left( \frac{GM}{V^3} \right),
\end{equation}

\no
at $R_{\rm V}$, where $V$ is the kick velocity and $M$ the black hole mass. A unit time 
in code units thus translates to $t = 155$~yr for a black 
hole mass of $M = 10^6 \ M_\odot$ and a kick velocity 
$V = 300 \ {\rm km \ s}^{-1}$. The scaling for the energy dissipation 
rate is,
\begin{equation}
 L = \left( \frac{q}{q_{\rm code}} \right)
 \left( \frac{V^5}{G} \right) \times \frac{{\rm d}E}{{\rm d}t},
\end{equation}

\no
where ${{\rm d}E}/{{\rm d}t}$ is the implied dissipation rate in code units,
 $q = M_d / M$ is the actual mass ratio between disc and black
hole and $q_{\rm code}=6 \times 10^{-4}$ is a fiducial value that we
use when plotting the derived energy dissipation rate.  For the same
kick parameters given above and $q=q_{\rm code}$, a typical code luminosity of $10^{-4}$ then
corresponds to a physical luminosity of $3.6 \times 10^{40} \ {\rm erg
\ s}^{-1}$.  Note that the peak luminosity depends upon the
ratio of the disc mass to the black hole mass and -- very sensitively
-- on the kick velocity, but that there is no dependence upon the
black hole mass.

\subsection{Results for perpendicular kicks}
%%%%%%%%%%%%%%%%%%%%%%%%%%%%%%%%%%%%%%%%%%%%%%%%%%%%%%%%%%%%%%%%%%% 
\begin{figure*}
\psfig{figure=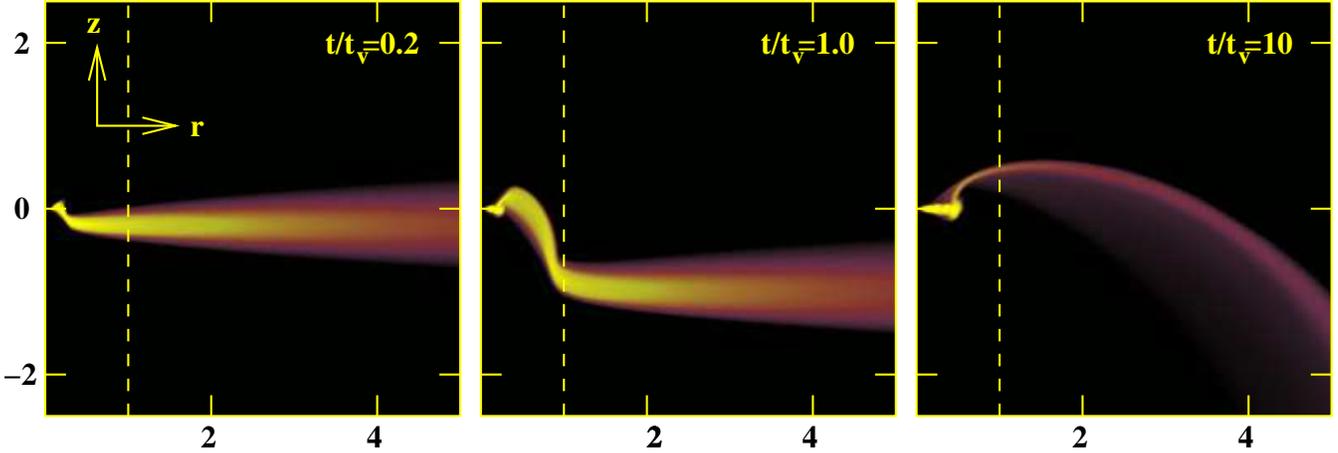,width=\textwidth}
\caption{Visualization of the density in the $(r,z)$ plane ($0 < r < 5$, $-2.5 < z < 2.5$) 
following a kick perpendicular ($\theta = 90^\circ$) to a disc with a surface 
density profile $\Sigma \propto r^{-3/2}$. By the end of the simulation the 
unbound material has been lost while the remaining bound disc has largely settled back 
into the equatorial ($z=0$) plane. The dashed vertical line in each image marks the  
cylindrical radius $r=1$.}
\label{figure_zeus90}
\end{figure*}
%%%%%%%%%%%%%%%%%%%%%%%%%%%%%%%%%%%%%%%%%%%%%%%%%%%%%%%%%%%%%%%%%%% 
%%%%%%%%%%%%%%%%%%%%%%%%%%%%%%%%%%%%%%%%%%%%%%%%%%%%%%%%%%%%%%%%%%% 
\begin{figure}
\psfig{figure=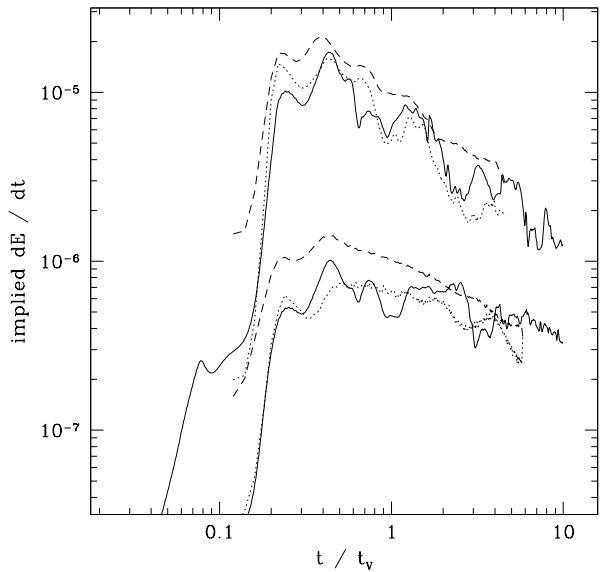,width=0.48\textwidth}
\caption{Comparison of the implied energy dissipation rate evaluated from ZEUS (solid curves) 
and SPH (long dashed curves) simulations for perpendicular kicks. The upper curves show results 
for steep surface density profiles ($p=3/2$), the lower curves results for $p=3/5$. The 
dotted curves show the portion of the implied SPH dissipation that arises from the quadratic 
term in the artificial viscosity.}
\label{figure_sph_zeus}
\end{figure}
%%%%%%%%%%%%%%%%%%%%%%%%%%%%%%%%%%%%%%%%%%%%%%%%%%%%%%%%%%%%%%%%%%% 

To model perpendicular kicks we use ZEUS in its axisymmetric spherical 
polar mode. For these runs the computational grid extends from 
$r = 0.025$ to $r=12$, and $0 < \theta < \pi$. The disc is set up in vertical
hydrostatic equilibrium with the mid-plane in the equatorial plane. 
The disc extends from $r=0.1$ to $r=10$. All luminosities are 
scaled to represent a disc mass ratio $q_{\rm code} = 6 \times 10^{-4}$. We use 
200 mesh points logarithmically spaced in $r$, and 200 mesh points in $\theta$.

Fig.~\ref{figure_zeus90} shows the evolution of the density in the ($r,z$) plane 
for a disc with a surface density profile 
$\Sigma \propto r^{-3/2}$. There is a clear distinction between the behaviour of 
the gas in the bound and unbound regions, which in this case are radially separated at $r=r_{\rm b}=r_{\rm ub}=1$. 
The bound gas is distorted by the kick 
into a bowl-like shape, which oscillates before settling back into the equatorial 
plane. The unbound gas departs on an essentially ballistic trajectory. A low density 
streamer of gas that is marginally bound persists out to quite late times.

The energy dissipation rate 
is plotted as the solid lines in Fig.~\ref{figure_sph_zeus}. The
luminosity peaks at about $t / t_{\rm V}=0.4$ for both of the surface
density profiles that we simulated ($p=3/2$ and $p=3/5$), before
declining slowly out to late times. The oscillations seen after the 
peak are present in both the SPH and the ZEUS results and also in 
ZEUS simulations run at different resolutions. Thus they appear to be 
physical. Since they do not appear for other kick angles they are 
probably associated with the dissipation of vertical shear in the 
$90^\circ$ case.

\subsubsection{Numerical tests}
Fig.~\ref{figure_sph_zeus} also shows a comparison between the ZEUS
light curve and that derived from the corresponding SPH simulation of
the same system. The SPH simulation -- which we discuss in more detail
in \S\ref{sec_out_of_plane} -- is fully three-dimensional (and hence not
optimal for modelling an axisymmetric problem) but is otherwise set up
and analysed in a manner that is as close as possible to the ZEUS
run. There are two main differences. First, individual SPH particles
(whose smoothing length can vary with time) have a fixed internal
energy, whereas in the ZEUS runs the sound speed is fixed in Eulerian
co-ordinates. Second, the SPH viscosity used for computing the
implied dissipation is similar but not identical to that used in ZEUS.
In particular, it includes both a linear and a quadratic term in the
divergence of the velocity field (see eq.~\ref{eq:q_sph} and eq.~\ref{eq:q_zeus}).
 Despite these differences we find good
agreement between the light curves derived from the SPH (dashed lines) and ZEUS (solid lines) 
simulations. The agreement is both in the overall normalization and in the rate of decay past
the peak. The agreement is not, however, perfect. In order to
identify the source of the discrepancy we calculated the implied SPH
dissipation due solely to the quadratic term in the artificial
viscosity. The resulting light curve is also shown as the dotted line in
Fig.~\ref{figure_sph_zeus}: it is in significantly better agreement
with the results of the ZEUS runs. This leads us to conclude that the
different treatment of the artificial viscosity is the main
source of differences between the two codes predictions.

We have also tested how sensitive the results are to the magnitude 
of the artificial viscosity. The potential for unphysical numerical 
dissipation requires particularly careful consideration for the 
case of perpendicular kicks, since in this case the initial perturbation 
is an $m=0$ warp that gives rise to a wave \citep{lubow93,ferreira08}. 
How such a wave physically dissipates is not obvious. To test the 
numerical reliability of the derived light curves, we have run additional 
$\theta=90^\circ$ simulations that differ by a factor of two in 
linear resolution (ZEUS, not plotted) or in mass resolution (SPH, 
shown later in Fig.~\ref{fig:res_test}). The higher 
resolution runs have lower effective numerical viscosity than 
the lower resolution realizations. For both sets of runs, we 
find that the shape and magnitude of the light curve {\em near the 
peak} is independent of the strength of the numerical viscosity. 
On the other hand,
the very early-time behaviour of the ZEUS light curve is found to be resolution (and thus viscosity) dependent 
in the sense expected if the dissipation is a numerical phenomenon.\footnote{ 
Higher resolution reduces the luminosity by an amount that is roughly 
consistent with the reduction in artificial viscosity.} Our 
conclusion from these tests is that the very early rise of the 
luminosity cannot be accurately modeled without the use of a 
physical model for wave dissipation within the disc. The behaviour 
of the light curve near the peak, conversely, appears to be robust 
and well-captured by the code viscosity. We interpret this robustness 
as being due to the much more violent fluid motions associated with the 
large distortions seen in Fig.~\ref{figure_zeus90}.

\subsection{Results for in-plane kicks: razor-thin discs}
\label{sec_razor_thin}
In-plane kicks were also simulated using both ZEUS and SPH. It is 
important to realize at the outset that, unlike in the case of 
perpendicular kicks, 
the ZEUS and SPH simulations of in-plane 
kicks model {\em different physical systems}. Our ZEUS runs are 
strictly two-dimensional in $(r,\phi)$, and thus model a razor-thin 
disc with a vertical thickness that is both negligible and constant 
with radius. Our SPH runs, on the other hand, model the effect of 
the same kick on a three-dimensional disc whose scale height is  
non-zero and increasing with radius. Although the resulting motions 
are still predominantly confined to the $(r,\phi)$ plane, the 
three-dimensional structure of the disc allows gas from large 
radii to flow inward over the surface of the disc ballistically 
{\em without} forming a prompt shock as is inevitable in a 
strictly two-dimensional system. As we will show, this results 
in significant changes to the resulting light curve. Here  
we discuss the ZEUS results, deferring the SPH results to 
\S\ref{sec_out_of_plane}, where they are presented as part of  
the investigation of arbitrary kick angles. We emphasize that 
it is not obvious which model is closer 
to physical reality. A real disc is of course three-dimensional, 
as modeled with SPH, but it is also much 
thinner, at the radii of interest,
 than the disc simulated numerically and hence arguably 
closer to the ZEUS razor-thin limit.

For the ZEUS simulations of in-plane kicks the computational grid extends from 
$r = 0.02$ to $r = 12$, with the initial disc occupying the radial 
range $0.1 < r < 10$. For this case, $r_{\rm b} = 0.172$ and 
$r_{\rm ub} = 5.8$. 
We use 400 grid points in $r$, logarithmically 
spaced to give a radial resolution $r_{i+1} / r_i \simeq 1.016$ at 
all radii, and 300 grid points in $\phi$. All luminosities are 
scaled to represent a disc mass ratio $q_{\rm code} = 6 \times 10^{-4}$.

%%%%%%%%%%%%%%%%%%%%%%%%%%%%%%%%%%%%%%%%%%%%%%%%%%%%%%%%%%%%%%%%%%% 
\begin{figure*}
\psfig{figure=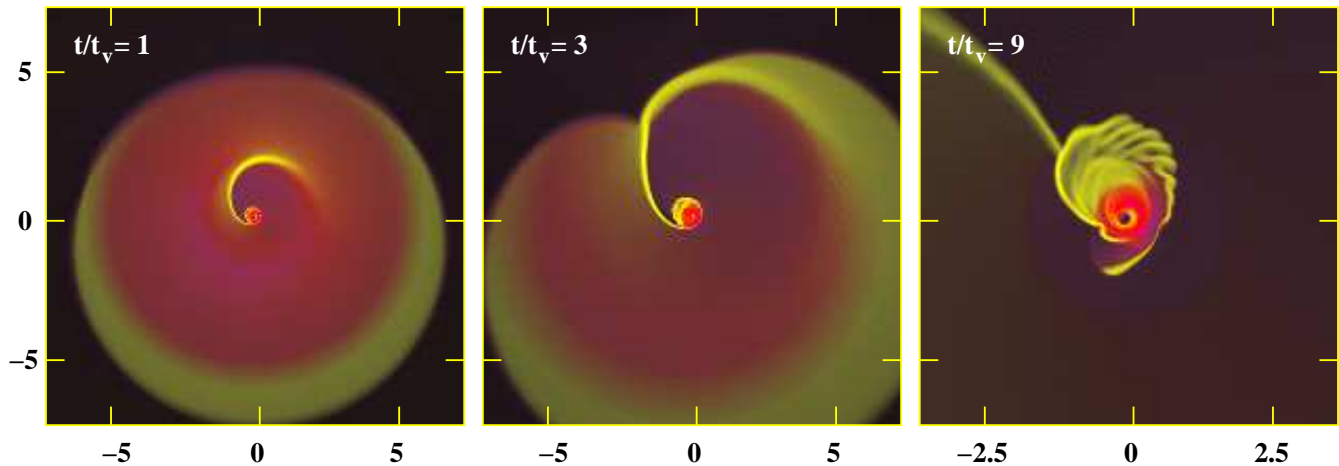,width=\textwidth}
\caption{Rendering showing the evolution of the surface density of the
disc following an in-plane kick. In this high resolution simulation
the disc extends up to $r=r_{\rm ub}$.  During the peak phase of
energy dissipation (left-hand panel) energy is dissipated at
successively larger disc radii as an outward moving wave propagates
through the gas. The outer part of the disc is unbound and escapes
ballistically. After the wave reaches the outer edge of the disc
(centre panel) the rate of decay of the energy dissipation rate
steepens markedly. At late times (right-hand panel, spatial and colour
scale adjusted to show structure in the innermost regions) low angular
momentum gas continues to accrete on to the bound remnant of the
original disc, releasing energy at a low level and forming a highly
non-axisymmetric accretion flow.}
\label{figure_zeus}
\end{figure*}
%%%%%%%%%%%%%%%%%%%%%%%%%%%%%%%%%%%%%%%%%%%%%%%%%%%%%%%%%%%%%%%%%%%

\begin{figure}
\psfig{figure=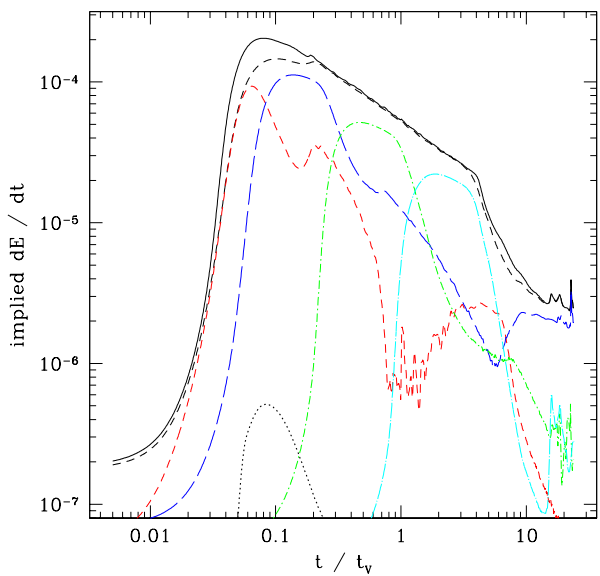,width=0.48\textwidth}
\caption{The implied energy dissipation rate as a function of time for
an in-plane kick into a razor-thin disc with a surface density profile
$\Sigma \propto r^{-3/2}$ and a fixed sound speed profile $c_s \propto
r^{-3/4}$. The disc extends up to $r=10$.  Time is measured
in units of the dynamical time ($t_{\rm V}$) at $R_{\rm V}$, while the
energy dissipation rate is in code units for a disc mass $M_{\rm d} /
M = 6 \times 10^{-4}$ between $r =0.1$ and $r=10$.  The solid 
black curve shows the total energy dissipation rate evaluated across
the whole computational domain, while the dashed black line is
the dissipation rate for $0.1 \leq r \leq 10$ for comparison with the SPH
results.  This latter has been divided up into individual contributions 
from annuli at successively greater radii: $0.1 < r < 0.172=r_{\rm b}$
(red short-dashed line), $0.172 < r < 0.55$ (blue long-dashed line), $0.55 < r < 1.79$ (green
dot-dashed line) and $1.79 < r < 5.8=r_{\rm ub}$ (cyan dot-long-dashed line). The dotted black 
curve (at the bottom) shows the result for a disc with no kick but an
instantaneous black hole mass loss of $0.03 M$.}
\label{figure_dedt_15}
\end{figure}

\begin{figure}
\psfig{figure=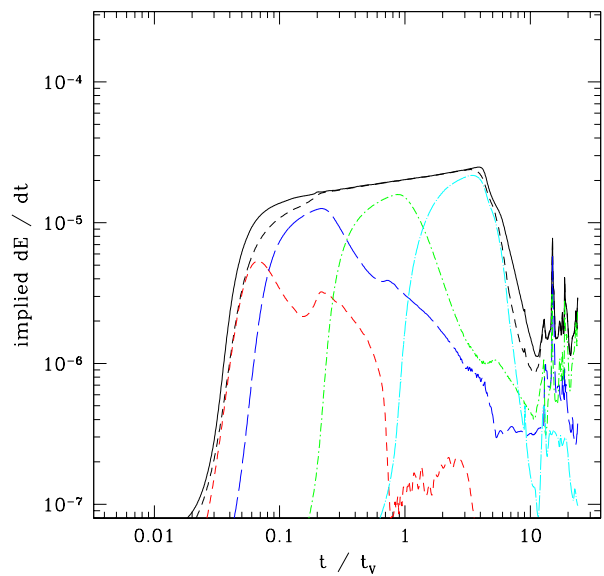,width=0.48\textwidth}
\caption{The implied energy dissipation rate as a function of time for a disc 
with a flatter surface density profile $\Sigma \propto r^{-3 / 5}$. The total 
energy dissipation rate (shown as the solid black curve) peaks at a 
smaller value than for $\Sigma \propto r^{-3 / 2}$. Light curves for 
individual annuli are plotted as in Fig.~\ref{figure_dedt_15}, except that no mass-loss light curve is shown here.}
\label{figure_dedt_06}
\end{figure}

The evolution of the surface density in the kicked disc is shown in 
Fig.~\ref{figure_zeus} for the case of a steeply declining surface 
density profile with $p=3/2$. We identify three main phases to the 
evolution, which can be matched to different parts of the bolometric 
light~curve plotted in Fig.~\ref{figure_dedt_15}. In the initial 
phase, a wave propagates outward through the disc, depositing energy 
in annuli at successively larger radii. The rate of energy deposition 
in the symmetric bound region of the disc ($r < r_{\rm b} = 0.172$), 
which includes both dissipation of the prompt energy input and 
accretion energy, peaks at 
$t / t_{\rm V}=  0.065$. For the $p=3/2$ model in which there is a substantial 
amount of mass at small radii, the overall luminosity peaks shortly 
afterwards ($t / t_{\rm V}=0.08$). The wave then moves on through the outer 
regions of the disc, during which time the luminosity decays 
roughly as a power-law with an index $s \simeq -0.6$. This phase 
ends at about the time when the wave reaches the outer edge of 
the disc. Afterwards, there 
is an intermediate period during which the total luminosity 
declines steeply. Finally, there is a phase in which the luminosity 
arises from the infall of low angular momentum gas into the 
inner regions of what remains of the disc. Inspection of animations 
of the simulation shows that the infalling gas shocks and enters 
the disc as dense streams. The inner disc at late times is 
highly non-axisymmetric and time variable. The associated 
dissipation rate is approximately constant but with large 
amplitude fluctuations (see also Fig.~\ref{figure_dedt_06}).

%%%%%%%%%%%%%%%%%%%%%%%%%%%%%%%%%%%%%%%%%%%%%%%%%%%%%%%%%%%%%%%%%%% 
%%%%%%%%%%%%%%%%%%%%%%%%%%%%%%%%%%%%%%%%%%%%%%%%%%%%%%%%%%%%%%%%%%% 
%%%%%%%%%%%%%%%%%%%%%%%%%%%%%%%%%%%%%%%%%%%%%%%%%%%%%%%%%%%%%%%%%%%
\begin{figure}
\psfig{figure=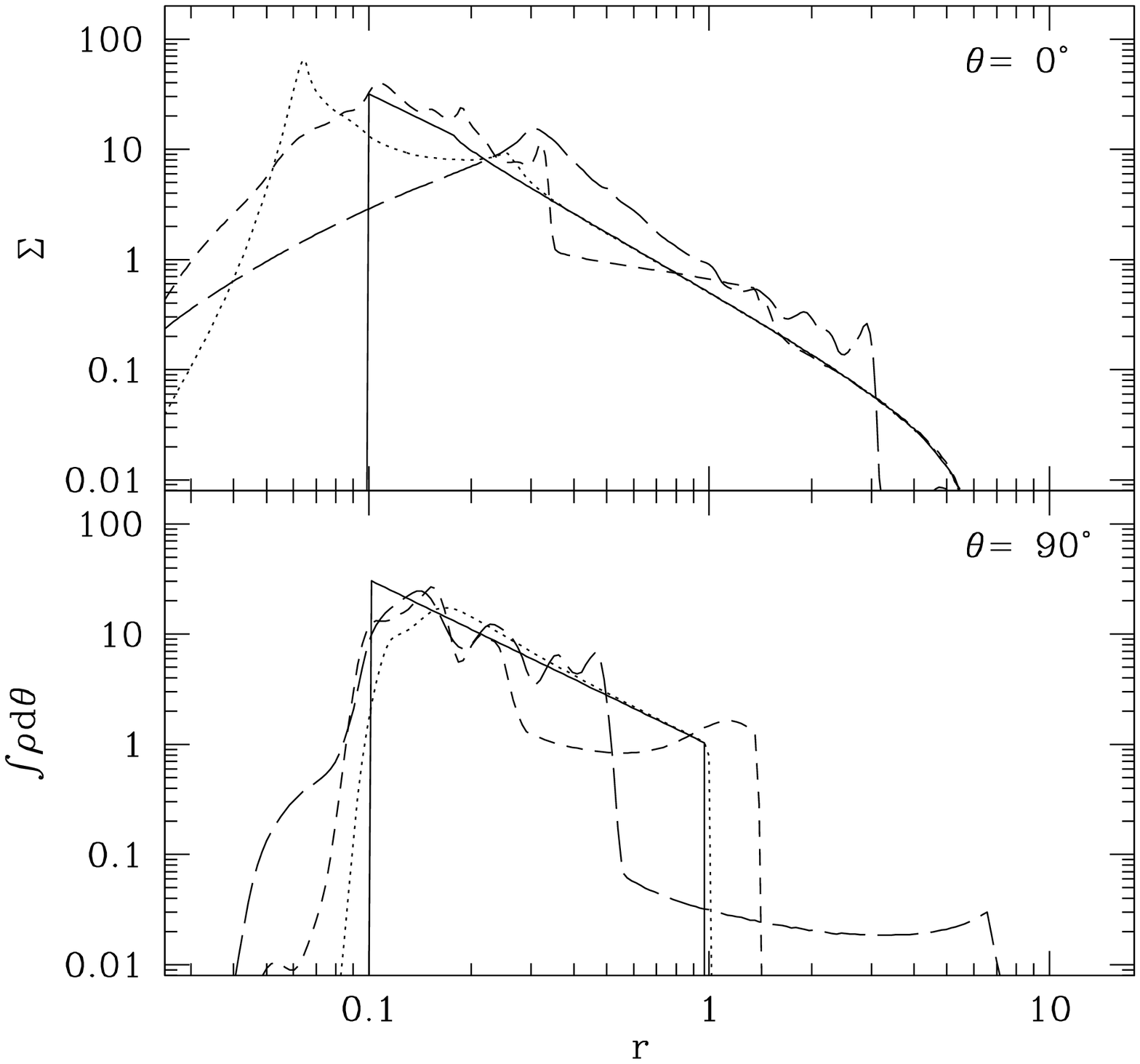,width=0.48\textwidth}
\caption{Evolution of the disc surface density with time from ZEUS
simulations of kicks at $\theta=0^\circ$ (upper panel) and
$\theta=90^\circ$ (lower panel) to the disc plane. For clarity only
gas that remains bound to the black hole is included in the
calculation of the surface density. Both discs had initial surface
density profiles $\Sigma \propto r^{-3/2}$. (solid lines, plotted here
in arbitrary units). Additional curves show the profile at $t/t_{\rm V}=0.1$
(dotted lines), $t/t_{\rm V} =1$ (short-dashed lines) and at the end of the
simulations (long-dashed lines plotted at $t/t_{\rm V}=20$ for the in-plane run
and $t/t_{\rm V}=10$ for the perpendicular kick).}
\label{figure_sigma_t}
\end{figure}
%%%%%%%%%%%%%%%%%%%%%%%%%%%%%%%%%%%%%%%%%%%%%%%%%%%%%%%%%%%%%%%%%%%

The same general behaviour is observed regardless of the choice of
initial surface density profile, though there are changes to the slope
of the light curve during the first phase and to the amount of low
angular momentum material that eventually
accretes. Fig.~\ref{figure_dedt_06} shows the energy dissipation rate
in a disc that has the same total mass but a shallower surface density
profile ($p=0.6$).  As expected, the amount of energy dissipated in
the outer annuli, whose contribution peaks at later times, is
increased relative to the $p=3/2$ disc model. As a result, the total
luminosity from the disc rises throughout the first phase, with a
power-law index $s \simeq 0.15$, until the wave leaves the disc. Since
the duration of the energy release is somewhat more extended the peak
luminosity is also reduced.

\subsection{Surface density evolution}
The different behaviour seen for in-plane kicks and perpendicular
kicks appears to reflect the increased importance of kick-induced
accretion for small kick angles. This is shown in
Fig.~\ref{figure_sigma_t}, where we plot the evolution of the surface
density of the bound portion of the disc for the in-plane
(upper panel) and perpendicular (bottom panel)
runs. For the perpendicular kick case we find that after
the kick the inner disc settles back to a surface density profile that
is very similar to that originally imposed. Note that this settling
process takes some time, and it is complete only out to $r \simeq 0.5$
by $t/t _{\rm V}=10$. A negligible amount of gas ends up interior to
the original inner edge of the disc. In contrast, there is substantial
accretion in the $\theta = 0^\circ$ run, and this is evident even at
early ($t/t _{\rm V}=0.1$) times when the light curve is undergoing
its initial rise.

\subsection{Mass-loss simulation}
\label{sec:massloss}
Fig.~\ref{figure_dedt_15} also shows how the energy dissipation rate
in the disc due to the kick compares to that which would arise as a
result of the black hole mass loss experienced upon merger (the dotted
line at the bottom of the figure).  For simplicity we set up a disc
model identical to that used for the kick simulation, but set $V=0$
and imposed an instantaneous mass loss of 3\%. The simulation was run
until $t/t_{\rm V}=2$. Given our parameters, the energy dissipation
rate in the disc induced by the mass loss peaks early ($t / t_{\rm
V}=0.085$) at a luminosity that is more than two orders of magnitude
below that generated by the kick. We can therefore consistently ignore
black hole mass loss in our simulations.  Of course this does not mean
that mass loss is always {\em physically} negligible, since (to quote
a trivial example) mass loss must occur in mergers whose symmetry is
such that no recoil is produced. The analytic discussion in
Section~\ref{sec:analytic_mass_loss} provides an estimate of when mass
loss might be competitive as a source of disc energy dissipation.

\subsection{Results for out-of-plane kicks}
\label{sec_out_of_plane}

%%%%%%%%%%%%%%%%%%%%%%%%%%%%%%%%%%%%%%%%%%%%%%%%%%%%%%%%%%%%%%%%%%%
\begin{figure}
\psfig{figure=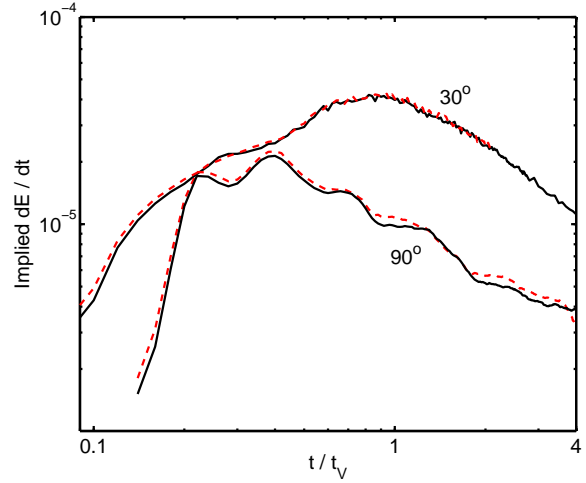,width=0.48\textwidth}
\caption[]{ Light curve obtained for $\theta = 30^{\circ}$ and $\theta = 90^{\circ}$, 
from SPH simulations at two different resolutions: $4$ million particle run (red dashed lines)  and
$8$ million particle run (solid black lines).}
\label{fig:res_test}
\end{figure}
%%%%%%%%%%%%%%%%%%%%%%%%%%%%%%%%%%%%%%%%%%%%%%%%%%%%%%%%%%%%%%%%%%%
%%%%%%%%%%%%%%%%%%%%%%%%%%%%%%%%%%%%%%%%%%%%%%%%%%%%%%%%%%%%%%%%%%%
\begin{figure*}
  \begin{center}
      \includegraphics[width=1.1\textwidth]{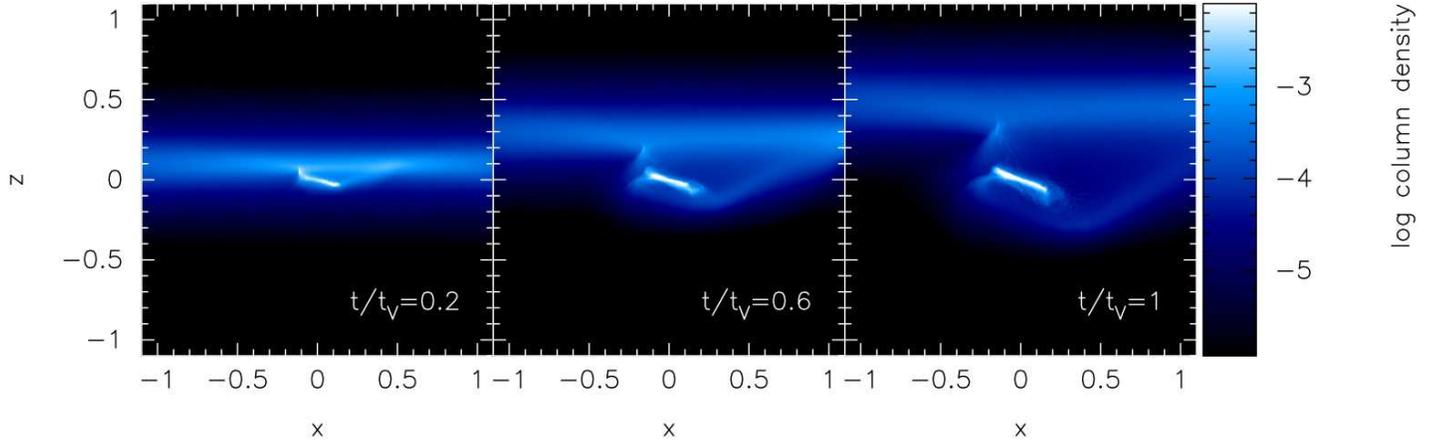}
      \caption[]{Visualization of the column density (in code units) at different times, 
from a simulation with a kick at $\theta =30^\circ$. Part of the disc remains 
bound to the black hole, and this material rapidly settles into a relatively stable 
tilted configuration. Unbound gas can be seen leaving the system in the upper 
portion of the panels.}
       \label{fig:edge}
  \end{center}
\end{figure*}
%%%%%%%%%%%%%%%%%%%%%%%%%%%%%%%%%%%%%%%%%%%%%%%%%%%%%%%%%%%%%%%%%%%
%%%%%%%%%%%%%%%%%%%%%%%%%%%%%%%%%%%%%%%%%%%%%%%%%%%%%%%%%%%%%%%%%%%
\begin{figure*}
   \begin{center}
      \includegraphics[width=1.1\textwidth]{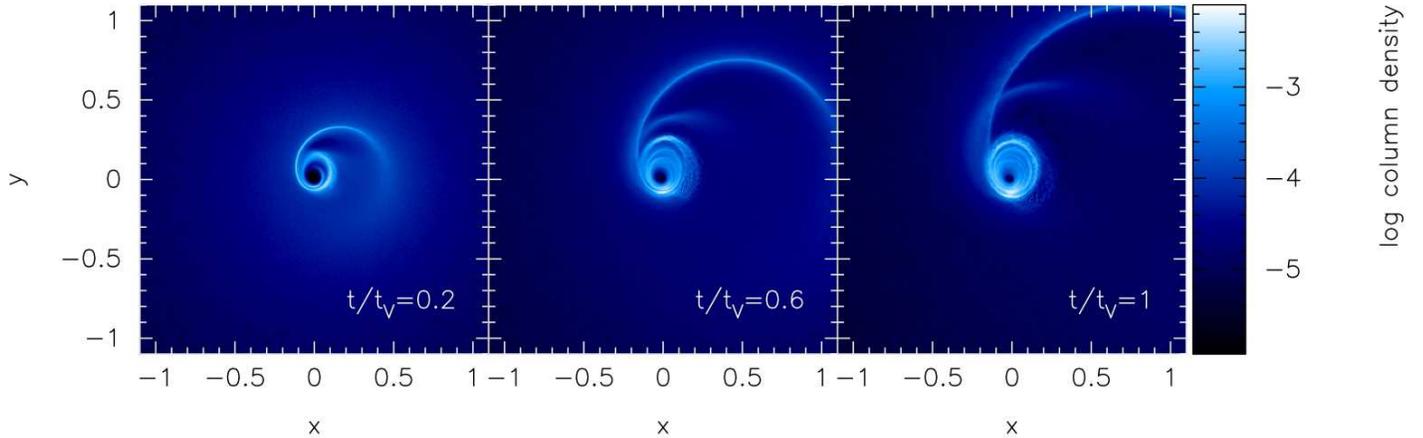}
      \caption[]{Visualization of the column density (in code units) through the disc at
different times, from a simulation with a kick at $\theta
=30^\circ$. In this face-on view, structure analogous to that seen in
the $\theta = 0^\circ$ simulations is clearly visible.}
      \label{fig:face}
   \end{center}
\end{figure*}
%%%%%%%%%%%%%%%%%%%%%%%%%%%%%%%%%%%%%%%%%%%%%%%%%%%%%%%%%%%%%%%%%%%

In this Section, we discuss the results of SPH simulations of discs
subject to kicks at arbitrary angles. Our simulations are for kick
angles $\theta = 0^\circ$ (with a finite thickness disc), $\theta =
15^\circ$, $\theta = 30^\circ$, $\theta = 60^\circ$ and $\theta =
90^\circ$. As previously, we adopt a nominal disc to black hole mass
ratio $q=6 \times 10^{-4}$. As an initial numerical check we show in
Fig.~\ref{fig:res_test} the light curves for $\theta=30^{\circ}$ and
$\theta=90^{\circ}$ at two different resolutions: 4 million particles
(dashed line) and 8 million particles (solid line).  Excellent
agreement is obtained, providing evidence that our simulations have
converged.  This result, together with the results of the code
comparison at $\theta=90^\circ$, makes us confident that any
uncertainties in the numerical computation of the light curves from
the SPH simulations are small. For practical purposes, they are certainly
much smaller than physical uncertainties arising, for example, from
poor knowledge of the actual initial surface density profile at
the relevant radii.

Visualisations of the evolution of the column density from the $\theta
= 30^\circ$ run are shown in Fig.~\ref{fig:edge} 
(edge-on view) and in Fig.~\ref{fig:face} (face-on view). 
For this kick angle the two characteristic radii 
are approximately equal to $r_b\approx 0.2$ and
$r_{\rm ub}\approx 4.8$, respectively.  Although the geometry is now more
complex the overall behaviour resembles that seen in the special cases
discussed earlier. As the black hole is ejected from the centre of the
disc it carries with it the innermost part of the disc, while the
outer part lags behind and is dispersed. The bound mass in the
simulation agrees with that predicted analytically and is constant
with time. The edge-on images show that there is a brief initial phase
during which the disc is warped, but as was found for $\theta =
90^\circ$ ,the gas quickly readjusts to orbit within a single plane. 
In this case, however, the plane is tilted with respect to that of the original
disc (see discussion in~\S\ref{sec:ang_mom}). 
The face-on images show that the final extent of the remnant
disc is comparable with $r_{\rm b}$. As
we will see in the following section, most of the dissipation occurs
in this region ($r\lsim r_{\rm b}$).

\begin{figure}
\psfig{figure=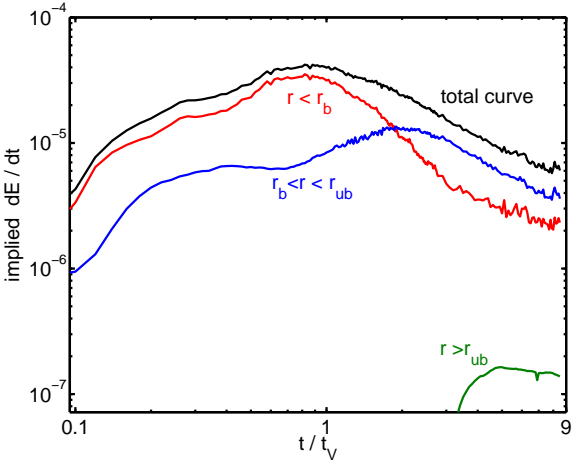,width=0.48\textwidth}
\caption[]{Light curve derived from an SPH simulation of the $\theta=30^{\circ}$ case.
${\rm d}E / {\rm d}t$ is in code units.
Here we also show the individual contributions arising from
energy dissipated in three different radial regions as marked.
Clearly most of the dissipation occurs within $R_{\rm ub}$.}
\label{fig:30deg_rings}
\end{figure}

The implied rate of energy dissipation as a function of time is
plotted in Fig.~\ref{fig:30deg_rings} with a black solid line.  The
contribution to the light curve from gas at different distances from
the black hole is shown in the same figure. We have divided the disc
into three broad annuli: $0.1<r < r_{\rm b}$, $r_{\rm b} < r < r_{\rm
ub}$ and $r > r_{\rm ub}$.  Within each annulus, we computed the
energy dissipation rate by summing over all particles that are {\em
instantaneously} within that region.  As was the case for the in-plane
kick, we see an initial rise in the luminosity as the compression wave
propagates outwards and energy is dissipated at successively larger
radii.  The peak in the total curve at $t \simeq 0.9$ is primarily due
to dissipation in the inner region ($r < r_{\rm b}$), although
material in the intermediate zone $r_{\rm b} < r < r_{\rm ub}$ --
whose contribution dominates the light curve during its declining
phase -- is responsible for a substantial amount of energy when
integrated over time. Within this ring ($r_{\rm b} < r < r_{\rm ub}$)
most of the dissipation occurs close to $r_{\rm b}$.  The contribution
to the light curve from gas at $r > r_{\rm ub}$ is negligible and
consistent with zero, as we assumed in the analytic
calculation\footnote{We verified in particular that particles {\it
originally} at $r>r_{\rm ub}$ are lost without encountering shocks or
compressions that result in significant heating.}.

\begin{figure}
\psfig{figure=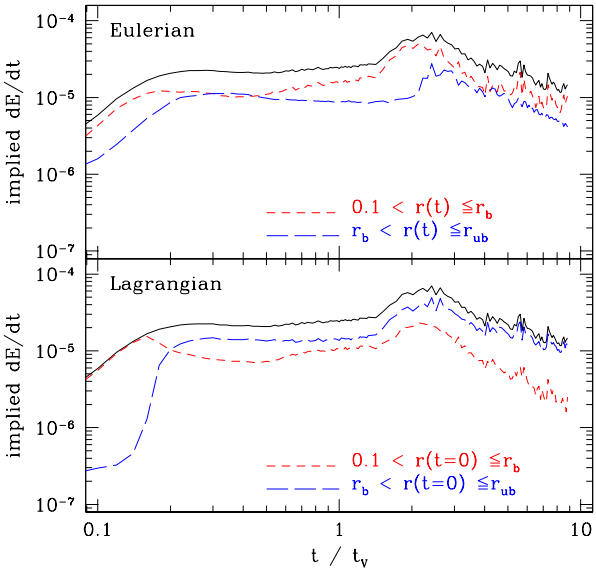,width=0.48\textwidth}
\caption[]{Light curve for $\theta = 15^{\circ}$ in code units.
This particular simulation used 4 million particles.
{\em Upper panel:} The total dissipation rate (black solid line) is the sum
of the energy dissipated within $r_{\rm b}$ (red short-dashed line) and
between $r_{\rm b}$ and $r_{\rm ub}$ (blue long-dashed line).
These contributions are calculated summing over particles present at that time 
in the region. {\em Lower panel:} the same as above, but here we show the individual 
contributions of those particles initially (at $t=0$) within $r_{\rm b}$  (red short-dashed line)
and between $r_{\rm b}$ and $r_{\rm ub}$ (blue long-dashed line).}
\label{fig:15deg_rings}
\end{figure}

We also note that both the total light curve and the light curves of
the two individual rings within $r_{\rm ub}$ show some evidence for a
two-component structure.  The presence of two components is more
clearly seen in the $\theta=15^\circ$ light curve, shown as
Fig.~\ref{fig:15deg_rings}.  For this figure we have computed the
dissipation not only as a function of the {\em current} radius of
particles within the disc (upper panel), but also by binning the
particles depending upon their {\em initial} location (lower
panel). This allows us to (at least partially) separate the
contributions from the prompt deposition of kick energy and that
arising due to accretion. We find that most of the dissipation occurs
within $r_{\rm b}$ (upper panel). At early times this dissipation is
partially associated with gas that originated in this region (lower
panel), and we thus infer that the initial rise (up to and through the
first shallow peak) in luminosity is due to both prompt dissipation
of kick energy and accretion that involves small changes in orbital
radius. The second more prominent peak, on the other hand, occurs as
energy is dissipated within $r_{\rm b}$ (upper panel) in gas that was
mostly originally in the region between $r_{\rm b}$ and $r_{\rm ub}$
(lower panel). We identify this contribution as being almost entirely
due to accretion of low angular momentum matter that falls from larger
radii and deposits energy at smaller radii.  The deposition occurs via
mild shocks and this may cause matter to lose additional angular
momentum, sink further toward the centre and therefore release
additional potential energy.
%%%%%%%%%%%%%%%%%%%%%%%%%%%%%%%%%%%%%%%%%%%%%%%%%%%%%%%%%%%%%%%%%%%
%%%%%%%%%%%%%%%%%%%%%%%%%%%%%%%%%%%%%%%%%%%%%%%%%%%%%%%%%%%%%%%%%%%

%%%%%%%%%%%%%%%%%%%%%%%%%%%%%%%%%%%%%%%%%%%%%%%%%%%%%%%%%%%%%%%%%%%
%%%%%%%%%%%%%%%%%%%%%%%%%%%%%%%%%%%%%%%%%%%%%%%%%%%%%%%%%%%%%%%%%%%

\begin{figure}
\psfig{figure=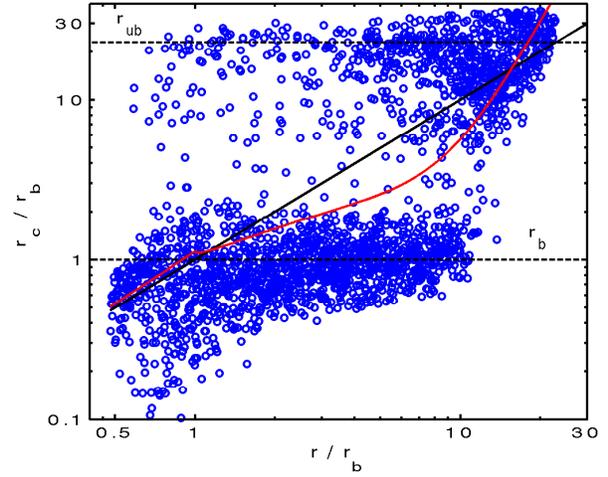,width=0.48\textwidth}
\caption[]{Final (at $t / t_{\rm V} =8.5$) radial position of a bound ($\epsilon <0$) SPH
particle in units of $r_{\rm b}$ as a function of its initial radial
position in the same units, for the $\theta = 30^{\circ}$ case (blue
symbols).  For clarity, only $10^{-3}$ of the total bound particles are
plotted. For comparison, we plot as the red solid line our analytic 
expectation (equation~\ref{eq:rc}).  The black solid line marks where
$r=r_{\rm c}$.}
\label{fig:rc_sim}
\end{figure}

In order to quantify whether accretion occurs beyond the level
expected from angular momentum conservation, we have made use of the
Lagrangian nature of SPH: we plot the final ($t / t_{\rm V} =8.5$)
radial position of bound particles as a function their initial position on a
particle-by-particle basis (Fig.~\ref{fig:rc_sim}). The data are from
the $\theta=30^\circ$ simulation.  In Fig.~\ref{fig:rc_sim}, the red
line shows the locus where particles should lie if they conserved their
angular momentum (equation~\ref{eq:rc}).  We find that to a {\em first
approximation} angular momentum conservation is a good assumption.
However, there is clearly a trend for the bound gas -- especially from
$r_{\rm b}<r<r_{\rm ub}$ -- to fall deeper into the potential well, and
form a remnant disc that is smaller than would be predicted under
our assumption. This is the result of angular momentum
mixing, where particles with larger initial angular momentum transfer
part of it to particles with an initially lower one.
 
\begin{figure}
\psfig{figure=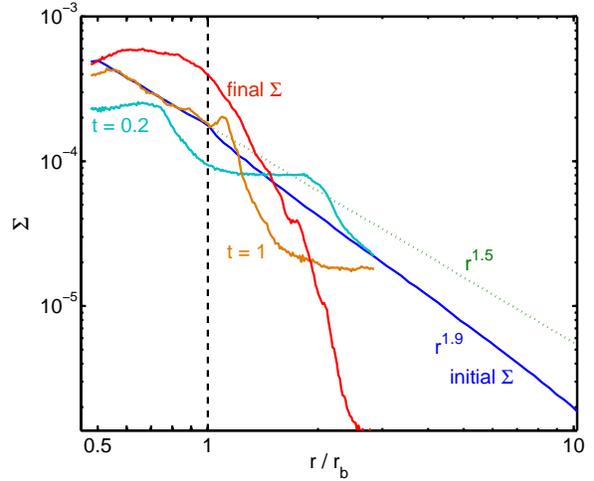,width=0.48\textwidth}
\caption[]{Surface density (code units) as a function of radius. At the beginning
of the simulation all particles are distributed with a $r^{-3/2}$
profile, while the bound particles follows a steeper profile of
$\Sigma \propto r^{-1.9}$. At the end of the simulation ($t / t_{\rm V}=8.5$) 
the bound particles are mostly within $R_{\rm b}$ and their surface density
distribution falls-off rapidly for larger radii.  }
\label{fig:sigma_teta}
\end{figure}

The final formation of a relatively compact disc can also be seen in
Fig.~\ref{fig:sigma_teta}, where we plot the evolution of the surface
density for this run. Most of the bound particles in the final disc
end up within $r_{\rm b}$, with a steep tail at larger radii that
drops to almost zero by about $3 r_{\rm b}$. This behaviour does not
match that predicted from the model assuming circularization at
constant angular momentum. Rather, there is an indication of angular
momentum transfer and additional accretion of matter (see also
Fig.~\ref{fig:rc_sim}). Since exact conservation of particle angular
momentum was assumed in our analytic model, these results imply that
it cannot provide a precise quantitative estimate of the total energy
release, but rather (for this kick angle) a lower limit.  Longer
duration simulations will be needed to determine how much energy in
excess of our simple model is indeed dissipated.

%%%%%%%%%%%%%%%%%%%%%%%%%%%%%%%%%%%%%%%%%%%%%%%%%%%%%%%%%%%%%%%%%%%
%%%%%%%%%%%%%%%%%%%%%%%%%%%%%%%%%%%%%%%%%%%%%%%%%%%%%%%%%%%%%%%%%%%

%%%%%%%%%%%%%%%%%%%%%%%%%%%%%%%%%%%%%%%%%%%%%%%%%%%%%%%%%%%%%%%%%%%
%\begin{figure}
%\psfig{figure=part_j_30deg.eps,width=0.48\textwidth}
%\caption[]{Final (at $t / t_{\rm V} =8.5$) specific angular momentum of a bound ($\epsilon <0$) SPH
%particle in code units as a function of its initial specific angular momentum,
%in the same units, for the $\theta = 30^{\circ}$ case (red
%marks).  For clarity, only $2 \%$ of the total bound particles are
%plotted. For comparison, we plot as black solid line where
%the marks should lie, for angular momentum conservation.}
%\label{fig:j_sim}
%\end{figure}
%%%%%%%%%%%%%%%%%%%%%%%%%%%%%%%%%%%%%%%%%%%%%%%%%%%%%%%%%%%%%%%%%%%

%%%%%%%%%%%%%%%%%%%%%%%%%%%%%%%%%%%%%%%%%%%%%%%%%%%%%%%%%%%%%%%%%%%
%%%%%%%%%%%%%%%%%%%%%%%%%%%%%%%%%%%%%%%%%%%%%%%%%%%%%%%%%%%%%%%%%%%

\begin{figure}
\psfig{figure=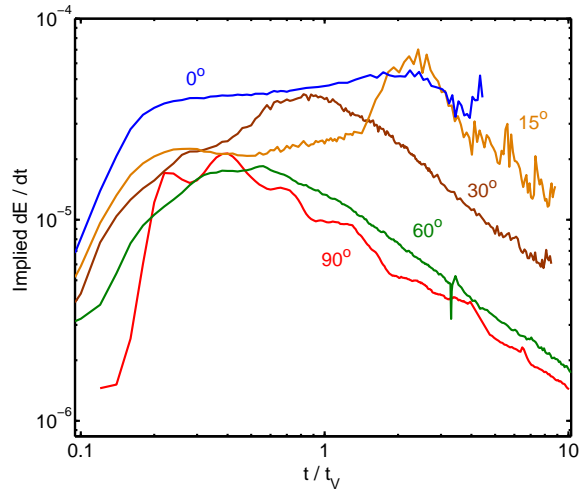,width=0.48\textwidth}
\caption[]{Total (from the whole disc) rate of energy dissipated as a
function of time for different recoil directions.  The figure shows
results from SPH simulations at $\theta=90^\circ$, $\theta=60^\circ$,
$\theta=30^\circ$, $\theta=15^\circ$, together with the in plane case
($\theta=0^\circ$).  All curves have the same disc model with a
surface density profile $\Sigma \propto r^{-3/2}$ and a total mass
(relative to the black hole) of $q=6 \times 10^{-4}$ between $r=0.1$
and $r=10$.  }
\label{fig:light_teta}
\end{figure}

The dependence of the shape and amplitude of the light curve on the
direction of the black hole recoil is illustrated in
Fig.~\ref{fig:light_teta}.  The peak luminosity and the total
dissipated energy increase for kicks that are more closely directed
toward the disc plane. This trend is evident for angles of $30^\circ$
and smaller (the $\theta=90^\circ$ and $\theta=60^\circ$ runs are
quite similar). The shape of the light curve is a complex function of
the kick angle. We interpret the behaviour seen in the simulations as
being a consequence of the existence of two components to the light
curve, whose separation in time is a function of kick angle. One
contribution is associated with gas in the inner disc (within and
close to $r_{\rm b}$). As the kick angle decreases $r_{\rm b}$ also
shrinks, and the energy dissipated within this region of the disc is
released on a shorter time scale. A second contribution, due to
accretion, occurs on a time scale that is related to $r_{\rm ub}$,
which {\em increases} for kicks that are directed closer to the disc
plane. The superposition of the two components results in a light
curve that is clearly double-peaked only for a narrow range of kick
angles, but whose evolution with kick angle qualitatively matches that
seen in the simulations.

 We also note that the result for the
$\theta=0^\circ$ simulation presented here, for a finite thickness
disc, differs significantly from that obtained with ZEUS for a razor-thin
disc \S\ref{sec_razor_thin} (compare
Fig.~\ref{fig:light_teta} with Fig.~\ref{figure_dedt_15}). In
particular, for the finite thickness disc we do not observe the very
rapid rise to peak luminosity seen in the ZEUS simulation.
Instead, the $\theta=0^\circ$ light curve continues the trend toward
longer time scale energy release that is seen in the $\theta=30^\circ$
and $\theta=15^\circ$ simulations. Analysis of the SPH
simulation suggests that the reason for the difference between the two
results lies in the presence of three-dimensional flows of gas across
the surface of the kicked disc. They cannot occur in a strictly
two-dimensional calculation. In the SPH runs the kick causes some of
the gas in the outer disc (at relatively large height above the
mid-plane) to flow inward over the surface of the disc, before
releasing energy at smaller radii and later times. Inspection of
the vertical profile of energy dissipation in the SPH run
shows a substantial release of energy in the surface layers of
the disc.

\begin{figure}
\psfig{figure=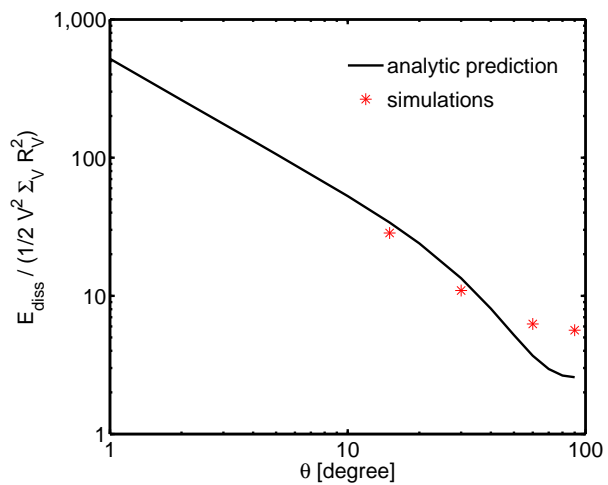,width=0.48\textwidth}
\caption[]{Comparison of the analytic prediction for the total energy 
release (solid curve) with the simulation results (stars), as a function 
of kick angle.}
\label{fig:ene_sim}
\end{figure}

Finally, we compare in Fig.~\ref{fig:ene_sim} the cumulative energy release 
as a function of $\theta$ with the analytic estimate discussed in 
Section~\ref{sec:analytic_mass_loss}. 
The figure shows (as the red stars) the total energy release 
derived by integrating the high and moderate $\theta$ light curves 
shown in Fig.~\ref{fig:light_teta} through to the end of the 
simulations. It is clear that energy dissipation is still ongoing at the 
(arbitrary) epoch at which the simulations stop, so our estimates 
of the total release are lower limits.  
Nonetheless, the simulations clearly show a trend toward increasing 
energy release for smaller kick angles that -- for kick angles between 
$90^\circ$ and $15^\circ$ -- agrees surprisingly well with that 
predicted analytically. For the $0^\circ$ run (whose integrated 
energy release is not plotted because the simulation was of 
shorter duration) the light curve (Fig.~\ref{fig:light_teta}) 
shows a further increase in total energy as compared to the 
$15^\circ$ run, though we do not find evidence for the very 
large (in fact divergent) increase predicted analytically. We 
attribute this difference as being due to the fact that the 
analytic divergence arises because a small amount of mass in 
the disc is predicted to circularize at zero 
radius. Any hydrodynamic mixing suffices to give the initially 
zero angular momentum gas a small amount of angular momentum, 
eliminating the divergent behaviour. This results in a 
smaller total energy release. We caution, however, that the maximum 
value of the total energy release at small kick angles may well 
depend upon the physical thickness of the disc, which is larger 
in the simulations than in real systems. For very thin 
discs the analytic trend toward increasing energy release might 
extend to smaller kick angles than those obtained here.

%%%%%%%%%%%%%%%%%%%%%%%%%%%%%%%%%%%%%%%%%%%%%%%%%%%%%%%%%%%%%%%%%%%
%%%%%%%%%%%%%%%%%%%%%%%%%%%%%%%%%%%%%%%%%%%%%%%%%%%%%%%%%%%%%%%%%%%

%%%%%%%%%%%%%%%%%%%%%%%%%%%%%%%%%%%%%%%%%%%%%%%%%%%%%%%%%%%%%%%%%%%
%%%%%%%%%%%%%%%%%%%%%%%%%%%%%%%%%%%%%%%%%%%%%%%%%%%%%%%%%%%%%%%%%%%

\section{Discussion}
\label{sec:discussion}

Our numerical results can be scaled to represent a wide range of
different systems. In doing so, three parameters enter the problem:
(1) the black hole mass, which affects the characteristic time scale
$t_V$ over which energy is deposited into the disc, (2) the kick
velocity, which affects both the time scale and the normalization of
the luminosity, and (3) the disc mass, which alters
the luminosity. The effect of kicks on surrounding discs might in
principle be observable in two distinct regimes. For {\em low-mass
black holes} with $M \sim 10^6 M_\odot$ the time scales for the
initial rise in the light curve in the source frame range from $\sim
10 \ {\rm yr}$ for $V = 300 \ {\rm km \ s}^{-1}$ down to as little as
$\sim 1 \ {\rm yr}$ for $V > 10^3 \ {\rm km \ s}^{-1}$.  It may
therefore be possible to detect a time-variable electromagnetic
counterpart from the kicked disc following a merger that has been
approximately localized by gravitational-wave (or other) observations
\citep{lippai08}.  For {\em high-mass black holes} ($M \sim 10^8
M_\odot$) the long time scales for variability and the rarity of
mergers in this mass range combine to make such triggered searches
unfeasible. It has been suggested, however, that the spectrum of a
kicked disc might be sufficiently distinct from that of other sources
as to allow identification in wide-area surveys (Schnittman \& Krolik
2008).

In addition to the black hole mass and kick velocity, the disc mass is
an important parameter, since the luminosity is proportional to the
surface density at the characteristic radius $R_V$. It is important to
note that this is a large radius: for $M = 10^6 M_\odot$ and $V = 300
\ {\rm km \ s}^{-1}$, we have $R_V = 0.05 \ {\rm pc}$ ($10^6 \ GM /
c^2$), while for $M = 10^8 M_\odot$ and $V=10^3 \ {\rm km \ s}^{-1}$
we have $R_V = 0.43 \ {\rm pc}$ ($\approx 10^5 \ GM / c^2$). At these
radii an argument can be made that the maximum surface density of a
geometrically thin accretion disc is limited by the onset of
fragmentation due to the disc's self-gravity
\citep{gammie01,goodman03,rice05,rafikov05,lodato07}.  A simple
estimate can be derived by making use of the steady-state models
calculated by \cite{levin07}. In these models the maximum value of
$\Sigma$ is a function only of $\Omega$, and varies from $\Sigma_{\rm
max} \sim 2 \times 10^{3} \ {\rm g \ cm}^{-2}$ at a radius where the
orbital period $P = 10^2 \ {\rm yr}$ down to $\Sigma_{\rm max} \sim 40
\ {\rm g \ cm}^{-2}$ or less at $P = 10^3 \ {\rm yr}$.  Writing the
maximum disc mass as $M_{\rm max} \sim \pi R_V^2 \Sigma (R_V)$, where
$\Sigma (R_V)$ is the maximum value of the surface density that would
be stable against fragmentation at $R_V$, we find that for a range of
kick velocities between $300 \ {\rm km \ s}^{-1}$ and $10^3 \ {\rm km
\ s}^{-1}$ the disc around a $10^6 M_\odot$ black hole would be
unlikely to exceed $q \sim 10^{-3}$ of the black hole mass. 
In scaling our results to $M =
10^6 M_\odot$, we therefore use the value $q=6 \times 10^{-4}$
discussed earlier.  For a $10^8 M_\odot$ black hole a somewhat more
massive disc could be stable, and for this case we scale to $q = 6
\times 10^{-3}$.

Given the important role that the assumed disc mass plays in determining the 
observability of emission from kicked discs, it is worth stressing that these 
estimates are crude. There are few observational constraints on discs at 
sub-pc scales in galactic nuclei. Theoretical work only strictly excludes 
the existence of {\em very} massive discs that are heated by viscous 
processes and cool radiatively. Therefore, it would not surprise us if our 
``limits" could be exceeded.
%%%%%%%%%%%%%%%%%%%%%%%%%%%%%%%%%%%%%%%%%%%%%%%%%%%%%%%%%%%%%%%%%%%
\begin{figure}
\psfig{figure=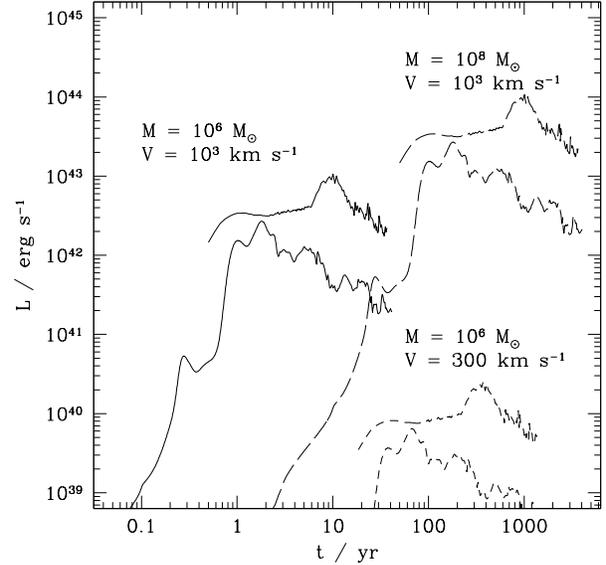,width=0.48\textwidth}
\caption{The predicted bolometric light curve of the kicked disc after scaling 
the numerical results to represent three different classes of systems. The 
solid curves show the light curves for $\theta=15^\circ$ (upper curve) and 
$\theta=90^\circ$  (lower curve) runs, following scaling to $M = 10^6 M_\odot$, $V = 10^3 \ {\rm km \ s}^{-1}$, 
and a disc mass ratio $q = 6 \times 10^{-4}$. Both runs are for a surface density profile 
with $p=3/2$. The short dashed curves are for 
identical parameters except for a lower kick velocity of $V = 300 \ {\rm km \ s}^{-1}$. 
The long dashed curves are for a system with $M = 10^8 M_\odot$, $V = 10^3 \ {\rm km \ s}^{-1}$, 
and a disc mass ratio $q = 6 \times 10^{-3}$.}
\label{figure_Lphysical}
\end{figure}
%%%%%%%%%%%%%%%%%%%%%%%%%%%%%%%%%%%%%%%%%%%%%%%%%%%%%%%%%%%%%%%%%%%

Fig.~\ref{figure_Lphysical} shows examples of the scaled light curves
for three sets of parameters: (1) a baseline case with $M = 10^6
M_\odot$, $V = 300 \ {\rm km \ s}^{-1}$ and $q = 6 \times 10^{-4}$;
(2) a more optimistic (as far as observability goes) case with $M =
10^6 M_\odot$, $V = 10^3 \ {\rm km \ s}^{-1}$ and $q = 6 \times
10^{-4}$; and (3) a high mass case with $M = 10^8 M_\odot$, $V = 10^3
\ {\rm km \ s}^{-1}$ and $q = 6 \times 10^{-3}$. These parameters are
similar to those considered by Schnittman \& Krolik (2008) except for
the fact that, for the reasons outlined above, we use a substantially
smaller disc mass.  For each parameter set we plot the total
luminosity from the $\theta = 15^\circ$ and $\theta = 90^\circ$
simulations, which roughly bracket the range of likely behaviour seen
in our simulations.

The maximum luminosity obtained from the $M = 10^6 M_\odot$ models is
$L \simeq 10^{43} \ {\rm erg \ s}^{-1}$, which corresponds to about
10\% of the Eddington luminosity $L_{\rm Edd} = 1.3 \times 10^{44} \
{\rm erg \ s}^{-1}$ for this mass of black hole. Reaching this
luminosity, however, requires a combination of circumstances that may
be uncommon: a high velocity kick ($V = 10^3 \ {\rm km \ s}^{-1}$),
directed almost into the plane of a disc, with a steep surface density
profile. The peak luminosity for most of the other models plotted in
Fig.~\ref{fig:light_teta} is smaller: around $2 \times 10^{42} \ {\rm
erg \ s}^{-1}$ (1.5\% $L_{\rm Edd}$) for $V = 10^3 \ {\rm km \ s}^{-1}$
and $\theta = 90^\circ$, and approximately $10^{40} \ {\rm erg \
s}^{-1}$ ($10^{-4} L_{\rm Edd}$ ) for the models with $V = 300 \ {\rm
km \ s}^{-1}$. These values encompass the range of luminosities ($6.3
\times 10^{-4} L_{\rm Edd}$ to $1.6 \times 10^{-2} L_{\rm Edd}$)
estimated by \citet{lippai08}.

\begin{figure}
\psfig{figure=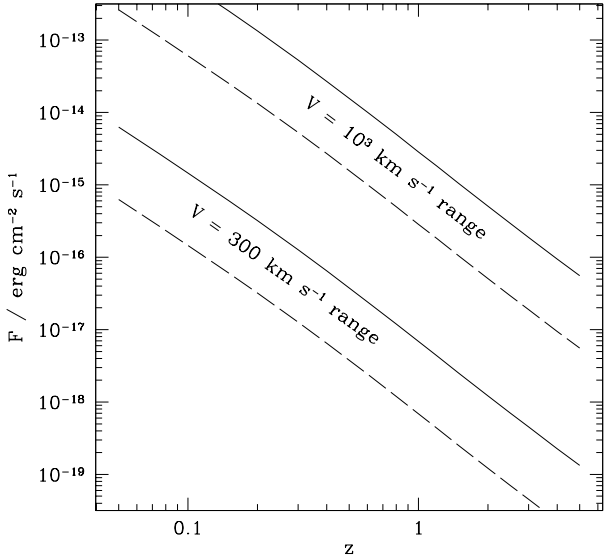,width=0.48\textwidth}
\caption{The predicted bolometric flux from the disc as a function of
the redshift $z$ of the source is plotted for two kick velocities,
$V=300 \ {\rm km \ s}^{-1}$ and $V=10^3 \ {\rm km \ s}^{-1}$. For each
velocity we show two curves that approximately encompass the range of
peak luminosities implied by our simulations. The lower dashed curves
show results for typical kick orientations (corresponding to
luminosities in code units of $10^{-5}$), while the upper solid curves
show results for more optimistic geometries in which the kick is
directed close to the disc plane ($10^{-4}$ in code units).}
\label{figure_flux_z}
\end{figure}

Fig.~\ref{figure_flux_z} shows the predicted bolometric flux from the
disc as a function of source redshift $z$. We assume a luminosity
distances appropriate for a standard cosmology. As expected from the
fact that even the most luminous sources are sub-Eddington, none of
our sources are predicted to be very bright. Electromagnetic
counterparts might be detectable if $V$ lies toward the upper end of
the considered range, especially if the merger is nearby. Low velocity
kicks ($V = 300 \ {\rm km \ s}^{-1}$), on the other hand, yield very
low luminosities that would not be detectable at plausible
cosmological distances.

Fig.~\ref{figure_Lphysical} also shows the predicted light curve for a
disc around a $10^8 M_\odot$ black hole following a kick of $10^3 \
{\rm km \ s}^{-1}$. These parameters have been chosen to match those
adopted by Schnittman \& Krolik (2008), although our curves are
calculated for a steeper surface density profile than the $\Sigma
\propto r^{-3/5}$ that they used. Nonetheless we find that (for the
$\theta = 15^\circ$ run) the time of the predicted peak ($t \sim 10^3
\ {\rm yr}$ in the source frame) is similar to that derived by
Schnittman \& Krolik (2008). The amplitude, however, is more than an
order of magnitude lower (about $10^{44} \ {\rm erg \ s}^{-1}$ as
opposed to a few $10^{45} \ {\rm erg \ s}^{-1}$). This difference
arises because we have limited the disc mass to the maximum that is
allowable before self-gravity would result in fragmentation. Unless
there is some way to circumvent this limit, we conclude that the
prospects for detecting the emission from kicked discs in surveys are
poor.

In addition to the energetic considerations discussed here, the actual 
observability of emission from kicked discs will also depend upon the 
spectral band in which the radiation is emitted. 
We expect emission in the soft X-ray, if the disc is 
optically thin at the radii of interest, or if the bulk of the energy 
is ultimately deposited in the upper layers of the disc atmosphere 
The emission would be at a temperature of $T_X \sim 1 \ {\rm keV}$, 
that corresponds to the post-shock value behind a shock of a few hundred 
${\rm km \ s}^{-1}$ \citep{lippai08,shields08}. Energy deposition at the 
midplane of an optically thick disc, conversely, would result in 
thermal emission in the infrared (Schnittman \& Krolik 2008). The 
results of \citet{levin07} imply that self-gravitating discs ought 
to be optically thick out to radii where the orbital period is 
$P \simeq 600 \ {\rm yr}$. Hence the high-velocity kicks that 
are most observable would deposit their energy at radii where the 
disc was optically thick. We caution, however, that this does not 
necessarily imply that all of the energy is thermalized into the 
infrared. Waves within discs can -- under certain circumstances -- 
deposit energy preferentially in the low density atmosphere rather 
than at the disc midplane \citep{bate02}. More detailed simulations, 
that include a realistic treatment of the disc's vertical structure, 
will be needed to determine where energy is deposited, 
and what the resulting spectral signatures are.

\section{Conclusions}
\label{sec:conclusion}
In this paper we have investigated the effect of a post-merger kick 
on the dynamics of a geometrically thin accretion disc, surrounding the 
newly merged black hole. The existence and strength of kicks following 
black hole mergers are now secure predictions of General Relativity. 
Whether small-scale gas discs typically attend mergers remains, however, 
uncertain. If gas discs are commonly present the perturbation to the 
gas caused by the kick will generate an electromagnetic counterpart 
to the merger, which may be detectable if its amplitude and time scale 
are observable at cosmological distances.

The main conclusions of our study are:
\begin{itemize}
\item[1.]
The dimensional assumption that the magnitude of energy release
is $(1/2) V^2 \Sigma_{\rm V} R^2_{\rm V}$  is an underestimate.
The energy available for dissipation varies strongly 
with the angle between the kick and the disc plane, and for kicks close to the plane
our analytic estimate exceeds $(1/2) V^2 \Sigma_{\rm V} R^2_{\rm V}$ by up to three 
orders of magnitude. A large increase in the energy release for kicks 
close to the plane is confirmed numerically.
\item[2.]
For most orientations of the kick, accretion energy (i.e. 
energy liberated when gas in the disc loses angular momentum and falls 
inward) dominates over the direct energy input to the gas in the 
frame of the kicked black hole. Accretion energy is particularly 
important for kicks that are directed toward the disc plane. The 
importance of accretion can be demonstrated analytically, and is 
confirmed by numerical simulations.
\item[3.]
We have run SPH and (for the special cases that are two-dimensional) 
finite difference numerical simulations to investigate the evolution 
of the disc and the rate of energy dissipation, following a kick at 
an arbitrary angle to the disc plane. The simulations 
yield explicit predictions for the form of the 
light curves as a function of both kick angle and surface density 
structure within the disc.
\item[4.]
The observability of emission from kicked discs depends upon the 
kick velocity, the orientation of the kick relative to the disc 
plane, and the mass of the disc at the radii where the energy is 
deposited. We have argued that the mass of the disc is limited by 
the requirement that the disc remains stable to fragmentation due 
to self-gravity. If this is true, the decreased disc mass (as 
compared to that assumed in prior works) outweighs the effect of the 
increased energy release per unit mass. As a result, the predicted 
luminosity 
around massive black holes is substantially smaller than previously 
thought. It is unlikely that such sources can be identified 
via wide-area sky surveys.
\item[5.]
The most feasible observational probe of the phenomena discussed 
here is via identification of variable disc emission following 
black hole mergers detected by other means (e.g. via detection 
of gravitational waves). Disc emission may be detectable provided 
that the kick velocity is large, especially in the case where 
the kick is directed close to the disc plane. 
To quantify, a kick velocity of $10^3 \ {\rm km \ s}^{-1}$ offers 
promising possibilities, but $300 \ {\rm km \ s}^{-1}$ appears 
to be very difficult or impossible to detect.

\end{itemize}
Physically, the magnitude and orientation of the kick is determined 
by the magnitude and direction of the spin of the black holes 
immediately prior to their merger. Our results emphasize that 
the observability of this class of electromagnetic counterparts 
depends sensitively on the distribution of the pre-merger spins. 
Studies, such as that by \citet{bogdanovic07} and that by \citet{king08}, that attempt to 
predict the evolution of black hole spin during the earlier phases of 
merger are thus particularly important for determining whether 
counterparts will be detectable. Determining the spectral signature 
of the kicked disc is also important, and this will require 
simulations of the disc evolution that include more complete 
treatments of the disc physics.

%--------------------------------------------------------------------------------------------------
%--------------------------------------------------------------------------------------------------

\section*{Acknowledgments} 
The simulations reported in this paper made use of computational facilities at 
the University of Leicester. Our visualizations made use of the {\sc SPLASH} 
software package \citep{price07b}. P.J.A. acknowledges support from
the NSF (AST-0807471), from NASA's Origins of Solar Systems program
(NNX09AB90G), and from NASA's Astrophysics Theory program (NNX07AH08G).
 
%----------------------------------------------------------------------------------------------
%-----------------------------------------------------------------------------------------------

\end{document}